\documentclass[a4paper,11pt]{article}
\usepackage{jcappub} 
\usepackage{lineno}

\usepackage{units}
\usepackage{url}
\usepackage{subfig}
\usepackage{graphicx}
\usepackage{caption}
\usepackage{hyperref}
\usepackage{amsmath}

\newcommand{\Offline}{\mbox{$\overline{\rm Off}$\hspace{.05em}\raisebox{.3ex}{$\underline{\rm line}$}}}

\arxivnumber{xxxx.xxxx} 
\title{POEMMA-Balloon with Radio: \\
       A multi-messenger, multi-detector balloon payload}

\author[hk]{J.~Adams,}
\author[hf]{J.~Alfaro}
\author[ba]{D.~Allard,}
\author[hj]{P.~Alldredge,}
\author[dp,dq]{R.~Aloisio,}
\author[dg]{R.~Ammendola,}
\author[da]{A.~Anastasio,}
\author[hl]{L.~Anchordoqui,}
\author[dg]{D.~Badoni,}
\author[ga]{J.~Baláž,}

% ...
\affiliation[ba]{Laboratoire APC, Université Paris Cité, Paris, France,}
% ...
\affiliation[da]{Sezione di Napoli, Istituto Nazionale di Fisica Nucleare, Naples, Italy,}
% ...
\affiliation[dg]{Sezione di Roma Tor Vergata, Istituto Nazionale di Fisica Nucleare, Roma, Italy,}
% ...
\affiliation[dp]{Laboratori Nazionali del Gran Sasso, Istituto Nazionale di Fisica Nucleare, L'Aquila, Italy,}
\affiliation[dq]{Gran Sasso Science Institute, L'Aquila, Italy,}
% ...
\affiliation[ga]{Institute of Experimental Physics, Slovak Academy of Sciences, Košice, Slovakia,}
% ...
\affiliation[hf]{Departments of Physics \& Astronomy and Astrophysics, Institute for Gravitation and the Cosmos, Pennsylvania State University, University Park, PA, United States,}
% ...
\affiliation[hj]{RSESC, University of Alabama Huntsville, Huntsville, AL, United States,}
\affiliation[hk]{CSPAR, University of Alabama Huntsville, Huntsville, AL, United States,} 
\affiliation[hl]{Department of Physics and Astronomy, Lehman CUNY, Bronx, NY, United States}
% ...

\emailAdd{jbe2130@columbia.edu}

\abstract{
A review of the current status of the field of Ultra‑High‑Energy Cosmic Ray (UHECR) including a summary of remaining open questions was 
presented in the white paper ``Ultra‑High Energy Cosmic Rays: at the Intersection of the Cosmic and Energy Frontiers" (Astropart. Phys. 147 (2023) 102794; arXiv:2205.05845). The authors concluded that two types of next‑generation detectors are needed to answer these questions: high‑accuracy instruments and detectors that maximize exposure at the highest energies. The Probe Of Extreme Multi‑Messenger Astrophysics (POEMMA), a proposed dual‑satellite observatory, exemplifies the latter class and is designed to increase statistics of the highest‑energy cosmic rays and to detect very‑high‑energy neutrinos following multi‑messenger alerts. POEMMA‑Balloon with Radio (PBR) implements a compact, balloon‑borne version of the POEMMA concept, adapted for a Super‑Pressure Balloon flight from Wanaka, New Zealand, with an expected campaign exceeding 20 days. PBR couples a wide field-of-view Schmidt telescope and a hybrid optical focal surface with a dedicated radio instrument to deliver simultaneous, complementary measurements of extensive air showers. The mission will validate the fluorescence detection strategy from space and raise technology readiness for a POEMMA‑like space mission by observing UHECR‑induced fluorescence light from suborbital altitudes, obtaining the first simultaneous optical Cherenkov and radio observations of high‑altitude horizontal air showers above the cosmic‑ray knee ($E>\unit[3]{PeV}$), enabling energy‑spectrum and composition studies at the PeV scale, and performing follow‑ups of multi‑messenger alerts to search for very‑high‑energy neutrinos via upward‑going air showers. This paper summarizes the PBR payload and its expected performance.
}

\begin{document}
\maketitle
\flushbottom


\section{Introduction}
\label{sec:intro}

Ground-based observatories such as the Pierre Auger Observatory \cite{PierreAuger:2015eyc} and the Telescope Array \cite{TelescopeArray:2008toq} have measured the spectra and composition of ultra‑high‑energy cosmic rays (UHECRs, $E\gtrsim1\,$EeV) with high precision and have begun to reveal arrival‑direction anisotropies~\cite{PierreAuger:2017pzq,TelescopeArray:2014tsd}. Despite these advances, the sources, cosmological evolution, and acceleration mechanisms of the highest‑energy particles remain unresolved. The community roadmap, developed in the context of the Snowmass 2021 decadal survey, emphasizes that answering these questions requires two complementary classes of next‑generation instruments -- detectors that deliver high‑precision shower and composition measurements, and detectors that maximize exposure at the highest energies \cite{Coleman:2022abf}.

Observing the atmosphere from above substantially increases exposure and enables near–full‑sky coverage. Satellite missions such as the proposed Probe Of Extreme Multi ‑ Messenger Astrophysics (POEMMA) and Multi-Messenger Extreme Universe Space Observatory (M-EUSO), the latest proposal of the Joint Exploratory Missions for Extreme Universe Space Observatory (JEM-EUSO) collaboration, follow this strategy to enhance sensitivity to both UHECRs and upward‑going air showers induced by neutrino interactions \cite{POEMMA:2020ykm,Plebaniak:2025ayq}. Stratospheric balloon payloads bridge ground-based and space-based observation: they allow critical subsystems to be exercised in a near-space environment and yield in-situ measurements that directly inform space-mission design.

The POEMMA‑Balloon with Radio (PBR), a part of JEM-EUSO, adopts this pathfinder approach. Building on the POEMMA concept and on prior balloon flights (notably EUSO‑SPB1 and EUSO‑SPB2)~\cite{Abdellaoui_2024, Adams:2025owi}, PBR combines a wide field-of-view Schmidt telescope (\autoref{subsec:optics}) with a hybrid optical focal surface (\autoref{subsec:FS}), a radio instrument (\autoref{subsec:RI}), and a $\gamma$/X-ray detector (\autoref{subsec:XGamma}) to obtain simultaneous complementary measurements of extensive air showers (EAS). The Fluorescence Camera (FC) images UV fluorescence with a 9216‑pixel multi‑anode PMT array sampled at 1\,$\mu$s; the Cherenkov Camera (CC) employs a 2048‑pixel SiPM design  with nanosecond‑scale sampling to capture fast Cherenkov pulses. The Radio Instrument (RI) is optimized for 60–500\,MHz and provides an independent measurement channel that complements the optical data.
The $\gamma$/X-ray detector monitors high-energy photon emission starting from \unit[15]{keV}, probing the early stage of shower development.

\begin{figure}[!h]
   \centering
   \includegraphics[width=\columnwidth]{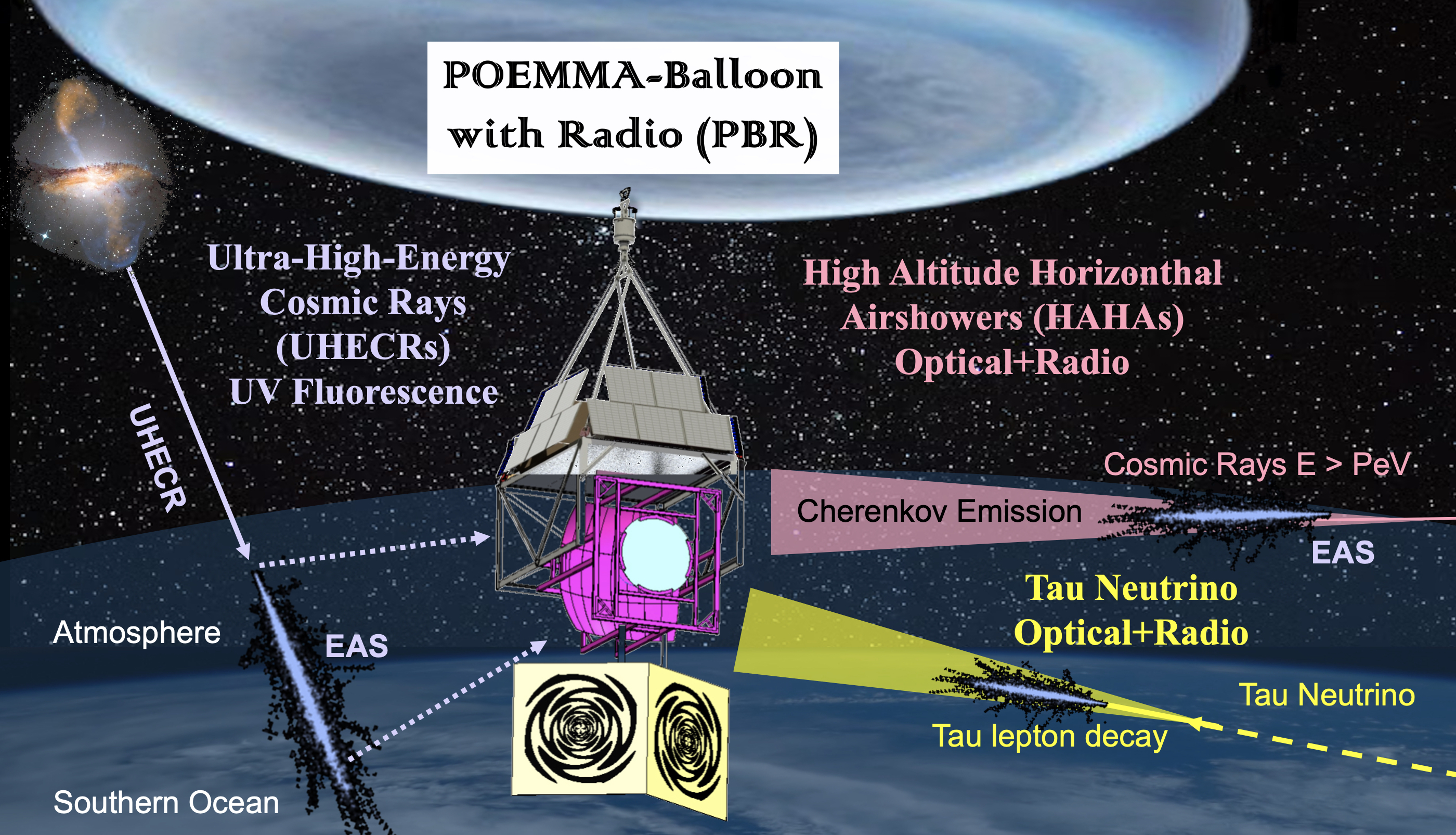}
   \caption{Main PBR scientific goals: observation of UHECRs via fluorescence from above; study a large set of HAHAs;  search for astrophysical neutrinos from multi-messenger targets of opportunity.}
	 \label{fig:science_sketch}
\end{figure}

PBR pursues the three main science goals sketched out in \autoref{fig:science_sketch}. First, PBR observes fluorescence from UHECR‑induced EASs at suborbital altitudes, demonstrating the fluorescence detection strategy planned for POEMMA‑like space missions and characterizing the performance of the relevant subsystems. Second, by recording a large sample of high‑altitude horizontal air showers (HAHAs), PBR measures the cosmic‑ray spectrum and composition near PeV energies and obtains the first simultaneous optical Cherenkov and radio observations of EASs at sub-orbital altitudes; a 20‑day super‑pressure balloon (SPB) flight at a nominal flight altitude of \unit[33]{km} from Wanaka, New Zealand is expected to increase the HAHA sample from less then 20 events, detected by EUSO-SPB2, to at least $10^{3}$ and enable composition studies near $\sim0.5\,$PeV. The suborbital altitude of PBR is crucial for these measurements, since the radio signal is difficult to measure using a POEMMA-like space mission due to ionosphere dispersion~\cite{Romero-Wolf:2013etm}, and signals from HAHAs do not reach ground-based instruments. Third, PBR searches for astrophysical neutrinos via targeted, rapid follow‑up of multi‑messenger targets of opportunity and by detecting upward‑going EASs produced when Earth‑skimming $\nu_{\tau}$ interactions yield $\tau$ leptons that decay in the atmosphere.

By integrating optical and radio channels on a single platform and target‑of‑opportunity operations, PBR provides unique scientific measurements that directly inform the design of future space-based observatories. The section that follows describes the three science goals (SG) in detail (\autoref{sec:science}); subsequent sections describe the instrument design~(\autoref{sec:instrument}), expected performance and sensitivity estimates (\autoref{sec:expPerformance}), before we summarize and conclude in~\autoref{sec:conclusion}.
\section{Science goals of PBR}
\label{sec:science}

\subsection{PBR SG1: UHECR observations from above}
\label{subsec:uhecr}

A primary cosmic ray entering the atmosphere initiates an EAS through repeated interactions with atmospheric nuclei. Secondary particles in the cascade excite atmospheric nitrogen molecules; the subsequent de-excitation produces isotropic UV fluorescence photons that track the longitudinal development of the shower. PBR will, for the first time, observe these fluorescence photons from UHECR-induced EASs at suborbital altitudes, thereby validating the detection strategy intended for future space missions such as POEMMA~\cite{POEMMA:2020ykm}.

The PBR Fluorescence Camera inherits its baseline design from the EUSO‑SPB1 and EUSO‑SPB2 fluorescence cameras \cite{Abdellaoui_2024,SPB2_FT}. EUSO‑SPB2 included upgraded, more sensitive optics designed to yield approximately one UHECR detection per 10 hours of observation (roughly two nights), according to simulations \cite{Filippatos:2021noz}. However, short flight durations limited the probability of detecting UHECR fluorescence events in those missions. PBR builds on the technological developments by increasing sensitivity and by introducing operational capabilities that were not previously available.

Unlike earlier flights, PBR can tilt its telescope to any elevation between nadir and horizontal. Tilting increases the geometric exposure to high‑energy events and thus improves statistics at the highest energies, while it also raises the effective energy threshold and modifies the detected event geometry. PBR will evaluate this pointing strategy on real data, quantify its impact on energy and shower‑maximum reconstruction, and thereby raise the technology readiness level (TRL) of relevant subsystems for a POEMMA‑like space mission (see \autoref{subsec_para:CC} for further telescope details).

\subsection{PBR SG2: High Altitude Horizontal Air-showers}
\label{subsec:hahas}

\begin{figure}[!htb]%
    \vspace{-5mm}
    \centering
    \subfloat {{\includegraphics[width=0.47 \linewidth]{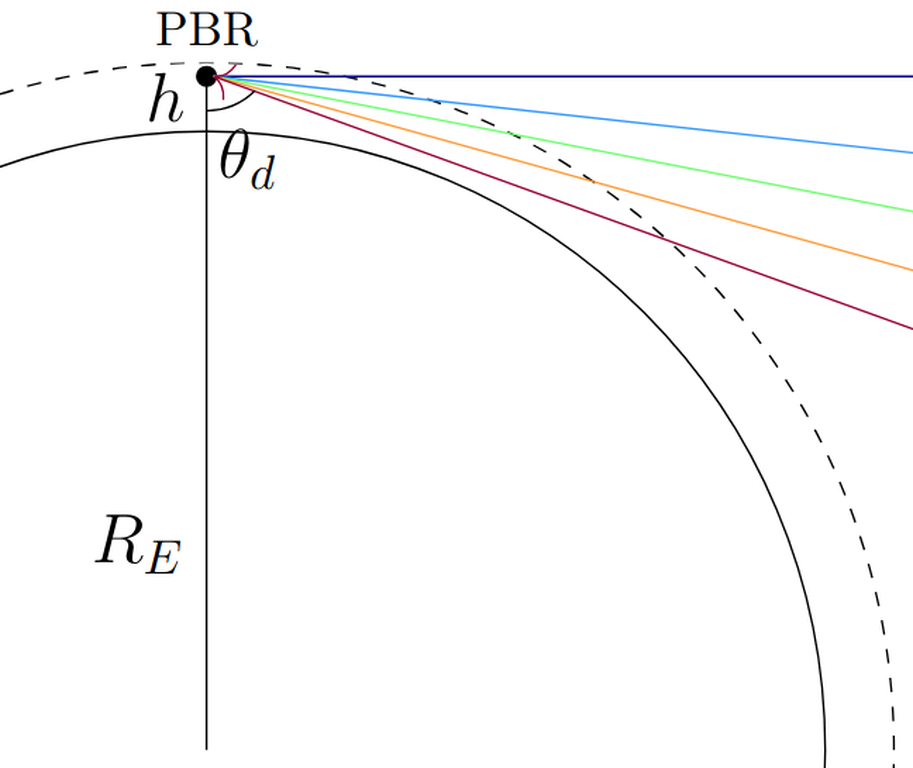} }}%
    \subfloat {{\includegraphics[width=0.53 \linewidth]{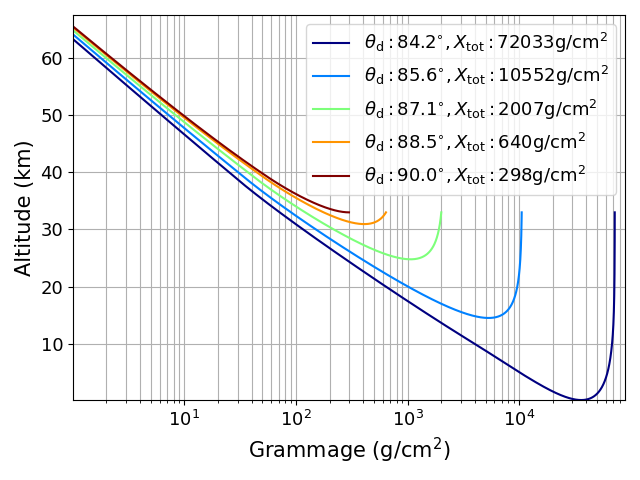} }}%
    \vspace{2mm}
    \caption{Left: HAHA trajectories with different inclination angles, $\theta_{d}$, relative to PBR nadir position at an altitude of $h$ above the Earth. $\theta_{E}$ corresponds to the Earth-centric angle of the HAHA trajectory, assuming the top of the atmsophere to be at $z_{\mathrm{atm}}$. Right: Altitude profile versus traversed atmospheric grammage for the $\theta_{d}$ as shown on the left.}%
    \label{fig:grammage_profiles}%
    \vspace{-5mm}
\end{figure}

HAHAs form the most frequent class of events that will be observed by the various detectors onboard PBR. HAHAs refer to EASs induced by cosmic rays that skim the Earth's atmosphere and traverse the telescope FoV, never intersecting the ground. The majority of HAHA shower development occurs above altitudes of 20~km, where the atmosphere is rarified, allowing for propagation over hundreds of kilometers. The geometry of an example HAHA event for a balloon-based detector and the corresponding altitude profile as a function of grammage  are shown in \autoref{fig:grammage_profiles}. For many trajectories through the atmosphere, a substantial amount of grammage is traversed, leading to nontrivial shower development. The extended path lengths combined with the unique atmospheric properties at high altitudes result in characteristic emission mechanisms that provide unique science potential for the PBR mission.

\paragraph{Optical Cherenkov Emission}
The propagating secondaries within a HAHA produce optical Cherenkov emission which can be detected by a high-altitude experiment (balloon-based, or satellite-based)~\cite{Cummings:2021bhg}. This forward boosted emission is produced in a narrow cone about the Cherenkov angle, which decreases from $\sim 1.4^{\circ}$ at sea level to less than a tenth of a degree for altitudes above 30~km, resulting in strong beaming. The threshold energy for optical Cherenkov emission for electrons in air increases from 22~MeV at sea level to greater than 200~MeV for altitudes above 30~km. The fraction of electrons exceeding this energy within the shower is a function of shower age, $s = 3X/(X+2X_{\mathrm{max}})$, where $X$ is the "slant depth" or integrated atmospheric density along the shower trajectory. At 30~km, the fraction of electrons above the Cherenkov threshold is only $15\%$ at shower maximum ($s = 1$), increasing to $>60\%$ for shower ages less than 0.2, allowing for improved sensitivity to early shower development compared to ground-based detection. Cherenkov emission from a HAHA event experiences minimal atmospheric attenuation, primarily through high altitude Ozone absorption, and is maximal for near-horizon trajectories that pass through low atmosphere.

\begin{figure}[!htb]%
    \centering
    \subfloat {{\includegraphics[width=0.49 \linewidth]{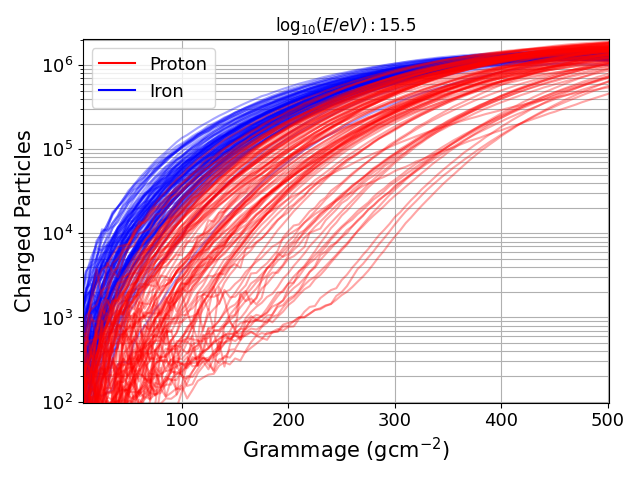} }}%
    \subfloat {{\includegraphics[width=0.49 \linewidth]{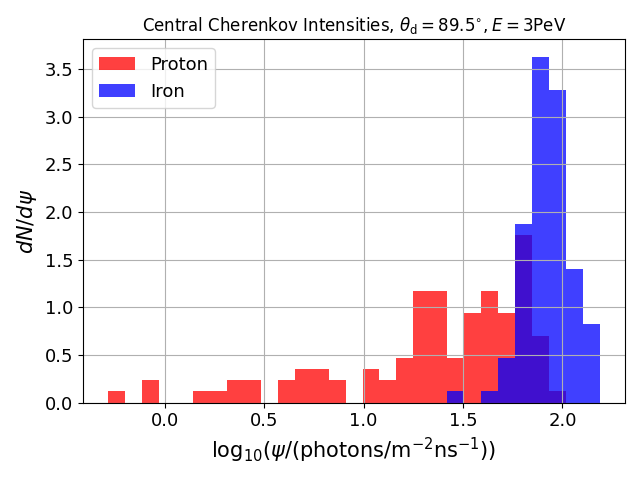} }}%
    \caption{Left: Early longitudinal development of 100 HAHAs induced by 3~PeV proton and iron primaries, as simulated with CORSIKA  (COsmic Ray Simulations for KAscade) \cite{Heck:1998vt}. Right: On-axis Cherenkov intensities $\Psi$ at payload for showers in the left panel with an arrival angle $\theta_{d} = 89.5^{\circ}$.}%
    \label{fig:development}%
\end{figure}

Measuring the early shower development of a HAHA event provides a means of measuring the chemical composition of the primary cosmic ray. The left panel of \autoref{fig:development} shows high altitude longitudinal development for EAS induced by 100 proton and iron primaries at 3~PeV up to an atmospheric grammage of $500 ~ \mathrm{g}/\mathrm{cm}^{2}$ ($\theta_{d} = 89^{\circ}$) while the right panel shows the maximum Cherenkov intensity (on-axis) for these showers as observed at 33~km, with an arrival angle of $\theta_{\mathrm{d}} = 89.5^{\circ}$. \autoref{fig:development} shows that EAS initiated by protons or iron are well separated in intensity at a specific slant depth (here 400~$\mathrm{g}/\mathrm{cm}^{2}$). Cherenkov angles above 20~km altitudes are less than $0.1^{\circ}$, resulting in well-reconstructed trajectories with minimal uncertainty on atmospheric absorption. The Cherenkov intensity from the early development of an EAS is highly sensitive to the primary composition, with EAS from iron primaries being brighter and having smaller variations. Using Reverse Monte Carlo or Energy Unfolding machine learning algorithms, the correlation between the observed cosmic ray rate as a function of energy provides an accurate method for determining the energy of cosmic rays. As discussed below in \autoref{subsec:expPerformance_HAHA}, low systematic uncertainties will allow for a measurement of the cosmic ray nuclear composition, with a peak sensitivity around 2~PeV (left panel of  \autoref{fig:pbr_events}). Thus, PBR will pioneer a new experimental methodology to address the composition/shape of the knee on the cosmic-ray spectrum.

\paragraph{Radio Emission}
The secondary electrons and positrons produced in a HAHA-type air showers are also expected to generate coherent radio emission through the geomagnetic effect--the separation of electrons and positrons from one another due to Earth's magnetic field~\cite{Huege:2016veh,Schroder:2016hrv}. This emission, as the optical Cherenkov one, is forward boosted. Above 30~km altitudes, the interaction length of electrons in air becomes comparable in scale to the gyroradius of electrons with energies exceeding hundreds of MeV, allowing for the effective separation of charged particles and the production of radio emission. 
Therefore, depending on the zenith angle and altitude of the shower development, we expect to be able to see a transition from two types of geomagnetic radio emission: dipole like emission caused by the induction of transverse currents and synchrotron-like emission caused by the circular deflection of the electrons and positrons~\cite{James:2022mea,Schluter:2022mhq,Chiche:2024yos}.


Radio emission from HAHAs has been observed by the Antarctic Impulsive Transient Antenna (ANITA) experiment, with 7 near-horizontal candidate events being detected during four long-duration flights  \cite{ANITA:2020gmv}.  Simulations of high-altitude radio emission estimate ANITA-IV's energy threshold to near-horizon class events to be roughly 1~EeV (using the two detected highly inclined air-shower candidates of ANITA-IV with viewing angles $\theta_{d} = 84.62^{\circ}$ and $84.36^{\circ}$) \cite{Tueros:2023lva}. In a similar manner, simulations of the Low-Frequency (LF) instrument for the Payload for Ultrahigh Energy Observations (PUEO) show an energy threshold of roughly 100~PeV to EAS induced by downgoing cosmic rays (after ice reflection) and neutrino-sourced $\tau$-leptons \cite{PUEO:2023saq}.
However, both ANITA and PUEO are self-triggered radio probes, and experience with ground-based air-shower arrays indicates that externally triggered radio detection can achieve a significantly lower detection threshold ($\sim$\unit[10s]{PeV}) \cite{Schroeder:2024bzf, Rehman:2023hw}. For this reason, the readout of PBR's radio antennas will be triggered by its air-Cherenkov instrument.

\paragraph{X- and Gamma-ray Emission}
\label{subsecpara:x/gamma}


The same mechanism that leads to coherent radio emission (gyration in the geomagnetic field of relativistic electrons and positrons in an EAS) also produces synchrotron radiation in the X/$\gamma$ band (photon energies $\gtrsim$\unit[10]{keV}) \cite{1972ZhPmR..16..452P,Saavedra:2025unx}. These high-energy photons are expected to be produced during the early stages of the shower development. Their detection would  provide a valuable constraint on the geometry and development of the air shower, contributing to a more accurate reconstruction of key parameters such as the primary particle energy and the depth of the first interaction, otherwise difficult to define with traditional detection techniques.
According to the Nerling shower development model \cite{NERLING2006421} the electron/positron component of young electromagnetic showers ($s \lesssim 0.2$) initiated by primary particles with energies $E \gtrsim 100~ \mathrm{PeV}$ produces secondary electrons above 1 TeV. Those high energy electrons produce a photon flux via synchrotron radiation and bremsstrahlung processes, spanning an energy range from a few tens of keV up to several tens of MeV \cite{Galper_2017, 2011NuPhS.215..250C}.
However, low-energy X-rays (in particular for $E \lesssim 30$ keV) are strongly attenuated over distances larger than a few kilometers. As an example, the mean free path of 30~keV photon is less than 2 km at 33 km altitude.
 The mean free path is calculated based on the mass attenuation factors from the NIST XCOM database~\cite{nist_xcom}.
 
The strong energy-dependent attenuation implies that high-energy photons measurements are only feasible with an {\it in situ } instrument, as X-rays and $\gamma$-rays in this energy range are largely absorbed in the atmosphere and remain inaccessible to both ground-based and space-based observatories.

Measuring these X‑rays with PBR would probe a relatively unexplored portion of the cascade and can improve reconstruction of primary energy and composition, while providing constraints on shower physics and interaction models.

\paragraph{Fluorescence Emission}
In addition to observing HAHAs via optical Cherenkov and radio emission, PBR will be able, for the first time, to measure the longitudinal profiles with the FC for some of these showers, mainly in tilt mode (see \autoref{subsec:uhecr}) depending on the shower geometry. Shower development in a rarefied atmosphere leads to a different average and spread of $X_{\mathrm{max}}$ (the depth at which the EAS reaches its greatest size in terms of secondary particles) compared to more vertical showers. This is a result of hadrons being more likely to decay than to interact for the case of more vertical showers. This unique shower development allows PBR to provide an alternative approach to evaluate hadronic interaction models. As these EASs develop significantly farther away from the FC than in the case of nadir observation, the energy threshold increases, and the detection rate is low.

\subsection{PBR SG3: Neutrino search from Targets of Opportunity}
\label{subsec:cc_ToO}

When pointed below the Earth's limb, PBR's sensitivity to the Cherenkov radiation (observed with the CC) and the radio emission (detected by the RI) from EAS makes it a pioneering instrument to study astrophysical sources of energetic neutrinos. With the Earth as a neutrino converter, high energy tau neutrinos can produce $\tau$-leptons that emerge from the Earth and initiate showers when they decay \cite{Venters:2019xwi,Cummings:2020ycz}. High energy muons from muon neutrino conversions in the Earth can also produce observable Cherenkov signals through EAS induced by stochastic losses in the atmosphere \cite{Cummings:2020ycz}.\\
While PBR’s small FoV and the short mission duration inherent to balloon missions severely limit its sensitivity to the diffuse cosmogenic neutrino flux even under the most optimistic UHECR source evolution models, the projected sensitivities of PBR (driven by the CC) to transient astrophysical neutrino point sources (through ToO observations) are comparable to current ground-based neutrino telescopes for bursts that occur during the mission, and will extend the neutrino energy range of sensitivity to beyond the PeV scale \cite{Reno:2021xos,Heibges:2023yhn,Wistrand:2023mpb,Posligua:2023cdm}. The radio instrument improves PBR's transient neutrino sensitivity above energies of 300~PeV, particularly considering its capability to make standalone daytime measurements \cite{PUEO:2023saq}.

The search for transient sources of very high-energy neutrinos (VHEN) is part of a multi-messenger program that includes multi-wavelength electromagnetic emission, cosmic rays, and gravitational waves \cite{Guepin:2022qpl,Ackermann:2022rqc}. The neutrino component of the multi-messenger program opened with IceCube's detection of a neutrino event in association with the flaring blazar TXS 0506+056 \cite{IceCube:2018dnn}. IceCube also found an excess of neutrino events associated with the nearby active galaxy NGC$\,$1068 at a $4.2\sigma$ significance \cite{IceCube:2022der}. 

High-energy neutrinos from astrophysical sources carry information about the acceleration of cosmic rays and their interactions that, together with observations of other messengers, can enable the extraction of characteristics of astrophysical environments (see, e.g., \cite{Guepin:2022qpl,Fang:2017tla,Keivani:2018rnh,Rodrigues:2023vbv}). 
Transients of interest for high-energy neutrino detection include blazar flares, tidal disruption events, binary neutron star mergers, gamma-ray bursts, X-ray binaries and supernovae. Alerts from the Gamma-ray Coordination Network (GCN) \cite{gcn}, the Transient Name Server (TNS) \cite{tns}, and Astronomer's Telegrams (ATel) \cite{atel} enable rapid follow-up for the PBR CC telescope \cite{Heibges:2025llt}. 
Steady sources such as TeV gamma-ray sources are also  of interest.

\subsection{Secondary Science}
The current design of PBR allows for multiple secondary science objectives without requiring additional hardware or observation plan changes compared to its primary objectives. 
\paragraph{Search for anomalous events}
The anomalous observation of steeply upgoing cosmic ray-like candidate events observed during the first and third ANITA flights~\cite{ANITA:2016vrp, ANITA:2018sgj} are of interest as their $27.4^\circ$ and $35^\circ$ elevation angles below horizontal make their existence difficult to explain within the standard model, due to large path lengths through the Earth. The data from IceCube~\cite{IceCube:2020gbx} and Auger~\cite{PierreAuger:2025hvl} have been used to conduct follow-up searches for these events, with neither finding a flux of similar showers above the expected background and, thereby setting stringent limits. However, these follow-ups have not been able to directly probe the energy nor the first interaction altitude ranges of the original ANITA observations. Rather, they have only been able to measure higher energies or first interaction points near or under the Earth's surface. Similarly, the fourth flight of ANITA observed 4 upward-going, near-horizon candidate events that are more consistent with standard model $\tau$-lepton emergence probabilities, but in strong tension with existing ground based detectors \cite{ANITA:2020gmv}. PBR is particularly well-suited to follow up the ANITA observations thanks to its position at the top of the atmosphere, horizon to nadir observation range, and its unique combination of fluorescence, Cherenkov, and radio measurements. These factors allow PBR to conduct the first multi-hybrid search for similar phenomena at energies as low as tens of PeV, fully covering the original ANITA observations' most probable energy, zenith, and first interaction height ranges. Therefore, with sufficient flight time, PBR may be able to give the relevant insights on these events. This follow-up will not interfere with the primary science goals as it can use any data taken during observation time in which the FoV of the telescopes includes directions at least $30^\circ$ below horizontal, which matches the observation time dedicated to the primary science goal of measuring UHECR showers from above via the FC.
\paragraph{Search for macroscopic dark matter through slowly evolving showers}
One possible characteristic of dark matter is that it is effectively dark due to having a high mass and, therefore, a low density in the universe.
This class of candidates, called macros, encompasses a range of phenomena, including nuclearites, strange quark nuggets, and primordial black holes \cite{Witten:1984rs, DeRujula:1984axn, SinghSidhu:2018oqs}.
The key feature of macros is that they would move at speeds well below the speed of light and would deposit significant amounts of energy in the atmosphere as they traverse it. 
This means they would leave an observable and clear signature in optical cosmic ray experiments if their trigger allows for the observation of slowly evolving phenomena.
The long readout time and high optical sensitivity of PBR enable it to observe or set limits on slow-moving objects like macros. 
A study addressing the capability of EUSO-SPB2 to make this measurement was conducted, and found significant possible exposure~\cite{Paul:2021bhh} depending on the nature of these hypothetical macros.
PBR is expected to have even higher exposure due to its wider FoV and lower energy thresholds.

\section{Instrument design of PBR}
\label{sec:instrument}

\begin{figure}[!ht]
\centering
    \includegraphics[width=.45\textwidth]{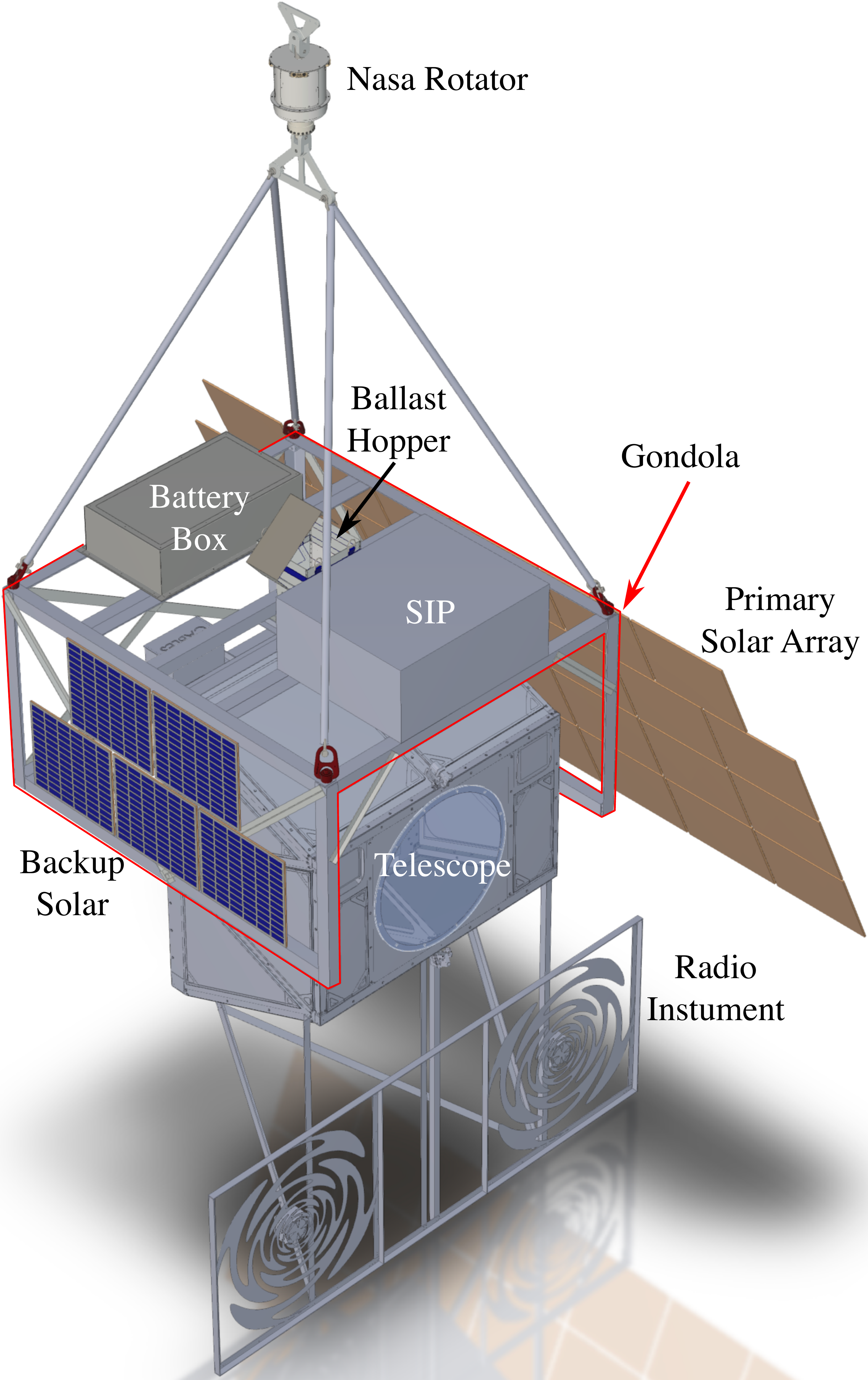}
    \caption{The PBR payload.}
    \label{fig:pbr-payload}
\end{figure}

\subsection{Gondola and Pointing}
\label{subsec:gondola_pointing}
The gondola is the main mechanical structure (highlighted in red in \autoref{fig:pbr-payload}) to support the science instruments of PBR.
It hosts all NASA‑required ballooning equipment, including the ballast hopper, Science Integration Packages (SIP), solar arrays (for both the science payload and NASA's Columbia Scientific Ballooning Facility -- CSBF), and the antenna booms which host all telemetry, GPS antennas as well as further equipment to monitor the balloon.
The gondola will be suspended from the NASA lightweight azimuth rotator, which supports efficient solar charging during the day and enables azimuth pointing for ToO follow‑up at night (see \autoref{subsec:cc_ToO}).
The SIP, science power system (see \autoref{subsec_para:PS}), and Gondola Control Computer (GCC), described in more detail in \autoref{subsec_para:GCC}, are mounted on the top of the gondola alongside the antenna boom, communication antennas, and ballast hopper.
The ballast hopper is at the gondola X-Y center of mass to simplify payload balancing and the ballast is routed around the telescope via tubing.

The telescope (including the Radio instrument) is hung beneath the gondola and can rotate in elevation angle on demand from $+15^\circ$ above horizontal to nadir via its tilting mechanism.
This capability enables the telescope to both point straight down to maximize exposure to UHECRs and also periodically point high enough to allow in‑situ checks of optical focus across the cameras' field of view (FoV) using stars.
The gondola has been designed so that the telescope's FoV is not obstructed at any point over its full elevation range, and the telescope rotates about its center of mass to minimize loads and torque requirements.
The tilting mechanism operates under a programmable logic controller similar to the model flown on EUSO‑SPB2, but with additional I/O lines for expanded sensor coverage (thermal probes, tilt sensors, photocells).
The flight tilt system has been fully tested in T/Vac at CSBF to minimize the risk of failure in the balloon environment. 

A primary design challenge was to ensure that the gondola and pointing systems meet the NASA flight safety requirements while keeping the overall design lightweight.
Details of the design are described below.

\subsection{Optical System}
\label{subsec:optics}

\begin{figure}[!htb]
    \centering
    \includegraphics[width=.75\textwidth]{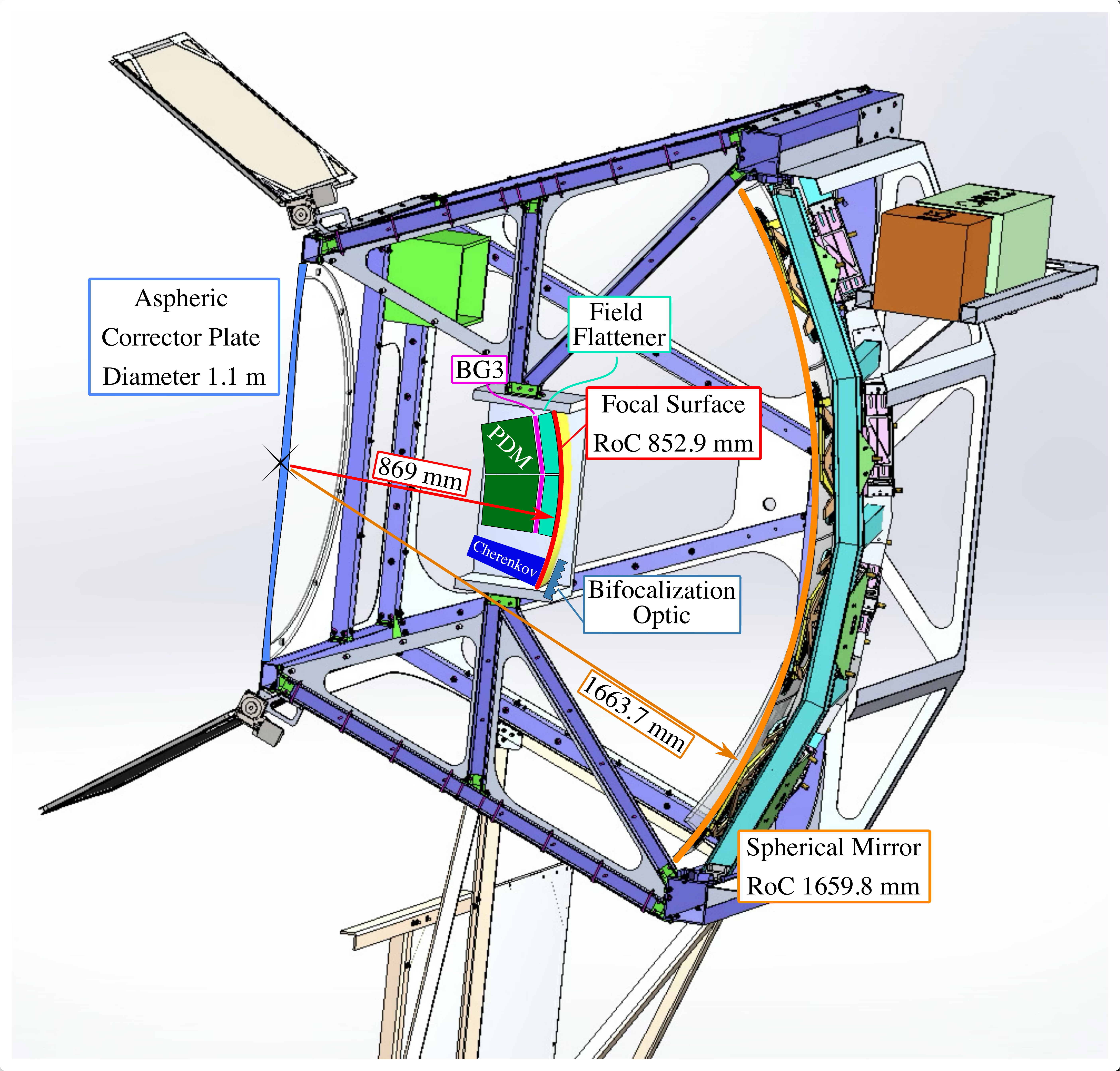}
    \caption{Optical sketch of the PBR modified Schmidt telescope, showing all major components.}
    \label{fig:optics_design}
\end{figure}

PBR adopts a modified Schmidt design (see \autoref{fig:optics_design}) with an entrance pupil of \unit[1.1]{m} diameter and a segmented, spherical primary mirror of approximately \unit[3.2]{m\textsuperscript{2}} (roughly \unit[1.82]{m}$\times$\unit[2.0]{m} maximum extent) and a radius of curvature of \unit[1.66]{m}. The mirror is build up from ten trapezoidal borosilicate segments (\unit[11]{mm} thick), produced by vacuum slumping, each coated with vacuum‑deposited aluminum and protected by a silicon‑dioxide overcoat. Two segment shapes suffice to span the 36$^\circ$$\times$30$^\circ$ FoV while avoiding unilluminated glass and hence excess mass. An aspheric PMMA corrector plate (ACP) at the entrance pupil corrects spherical aberration; it is diamond‑turned and polished to a central thickness of $\simeq$\unit[10]{mm} tapering to $\simeq$\unit[5]{mm} at the edges. 

\begin{figure}[!htb]
    \centering
    \subfloat {{\includegraphics[width=.75\textwidth]{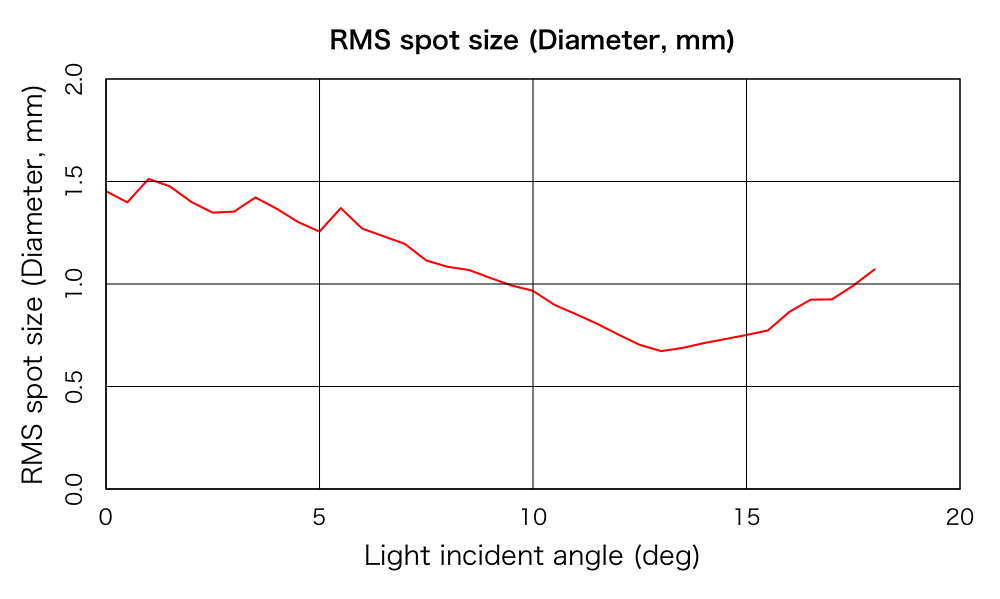} }} \\
    \subfloat {{\includegraphics[width=.75\textwidth]{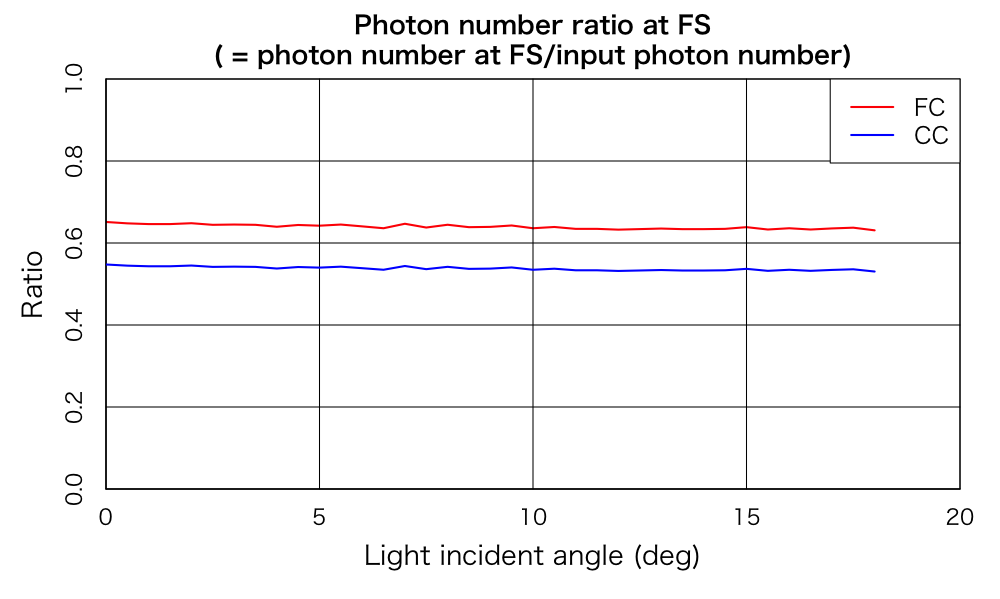} }}
    \caption{ PBR optical system results from ray tracing using Zemax OpticStudio designed parameters for photons at \unit[337]{nm}, \unit[357]{nm}, and \unit[391]{nm} (1:1:1). Top: root-mean-square spot size diameter for different light incident angles. Bottom: Throughput for each camera given as the ratio between injected and arriving photons throughout the FoV of the instrument.}
    \label{fig:optics_performance}
\end{figure}

We modeled the system in Zemax OpticStudio (Ansys) using photons at \unit[337, 357, 391]{nm} in equal proportion. The simulations indicate a point‑spread function (PSF) with RMS diameter of less than \unit[1.5]{mm} (top panel of \autoref{fig:optics_performance}), and encircled energy within one pixel exceeds 80\% across the expected incidence angles. We estimate the throughput to be above 65\% for the FC and above 55\% for the CC; the CC's lower throughput arises from losses in the Optical Accordion (OA), described below.

\begin{figure}[!htb]
    \centering
    \subfloat {{\includegraphics[width=.45\textwidth]{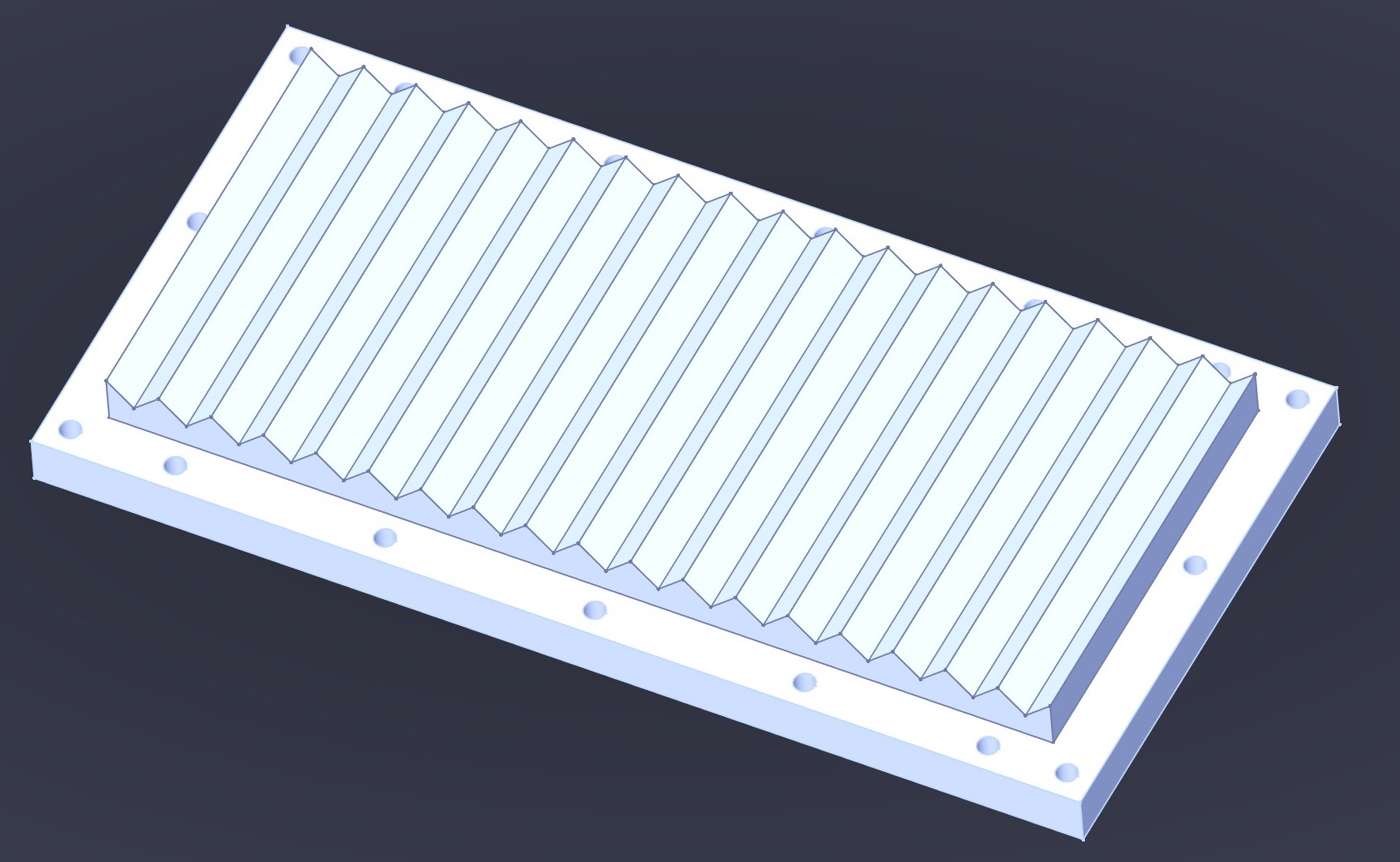} }}
        \hfill
    \subfloat {{\includegraphics[width=.45\textwidth]{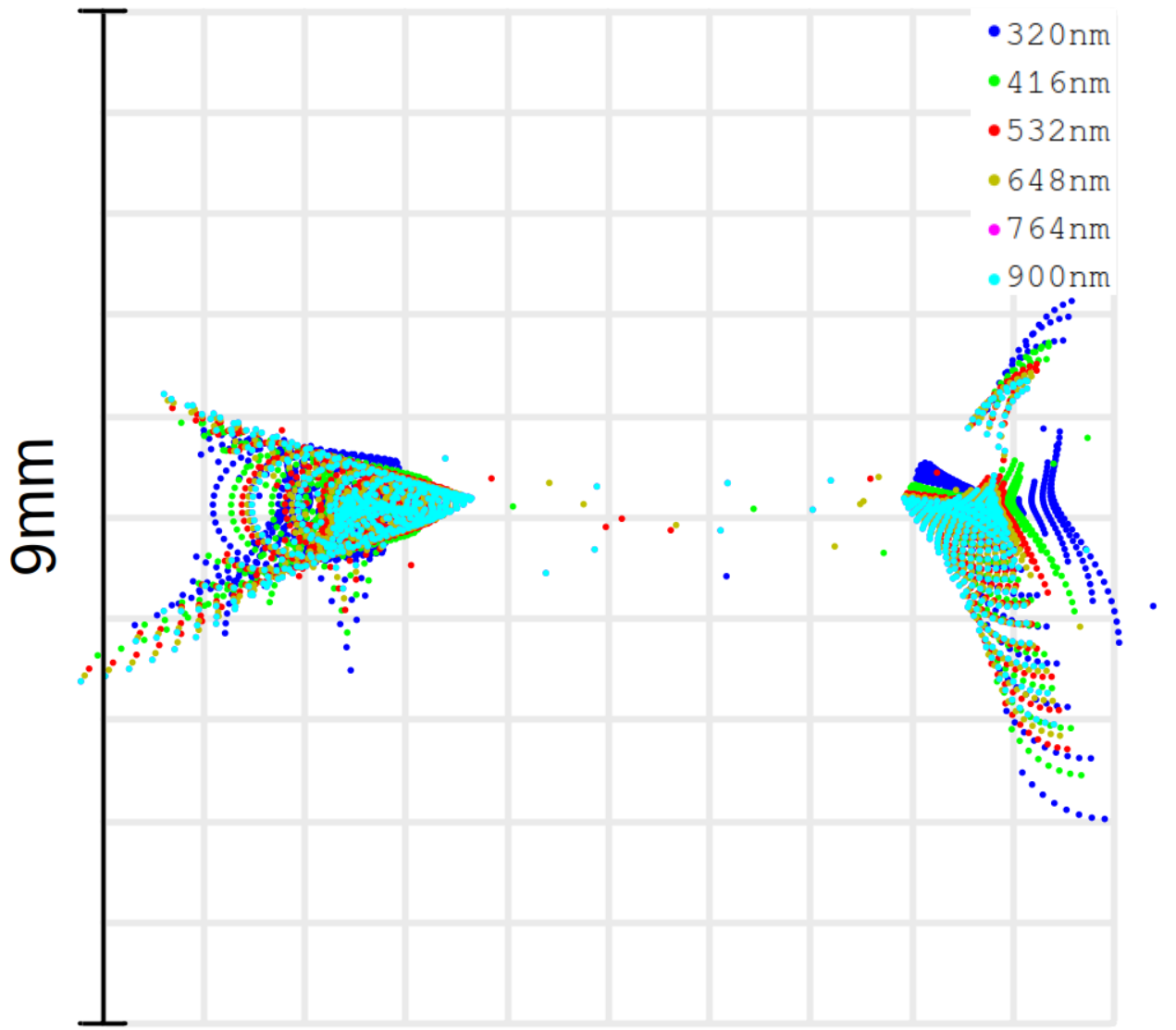} }}
    \caption{Left: CAD rendering of the OA showing the prism structure on one side, while the opposite is flat. Right: Spot diagram of the OA showing the desired \unit[6]{mm} separation with most energy contained in a single pixel for various wavelengths.}
    \label{fig:OA_performance}
\end{figure}

Each camera also incorporates optical elements to suppress background and improve image quality. We designed a PMMA Optical Accordion (OA) that, when placed \unit[30]{mm} in front of the CC, produces two image spots on the CC focal plane. Requiring time‑coincident detection of both spots suppresses noise triggers caused by direct cosmic‑ray hits on the CC. The OA has one flat face and one face carrying a one‑dimensional prism array (\unit[29]{mm} spacing), producing a spot separation of approximately two pixels. \autoref{fig:OA_performance} shows Zemax simulation results for the OA.

Because the FC PDMs are flat (unlike the CC's curved surface), we place a small field‑flattening lens in front of each FC PDM to preserve the PSF. Each PDM is also fitted with a BG3 filter to limit sensitivity to \unit[290-430]{nm} and thereby reduce optical background.

\subsection{Telescope Mechanical Structure}
\label{subsec:TelMech}
\begin{figure}[!h]
    \centering
    \begin{minipage}[t]{0.48\linewidth}
        \centering
        \includegraphics[height=6cm]{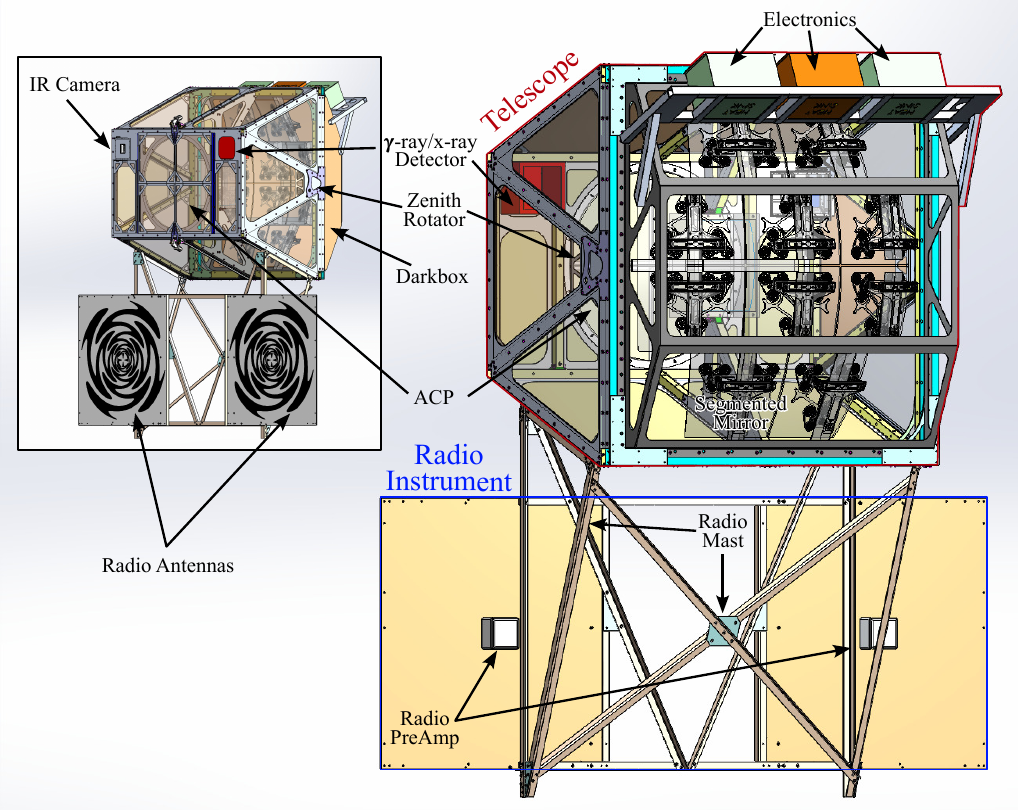}
        \caption{The PBR Telescope and Radio Assembly. Major components of both systems are labeled.}
        \label{fig:pbr-telescope}
    \end{minipage}%
    \hfill
    \begin{minipage}[t]{0.48\linewidth}
        \centering
        \includegraphics[height=6cm]{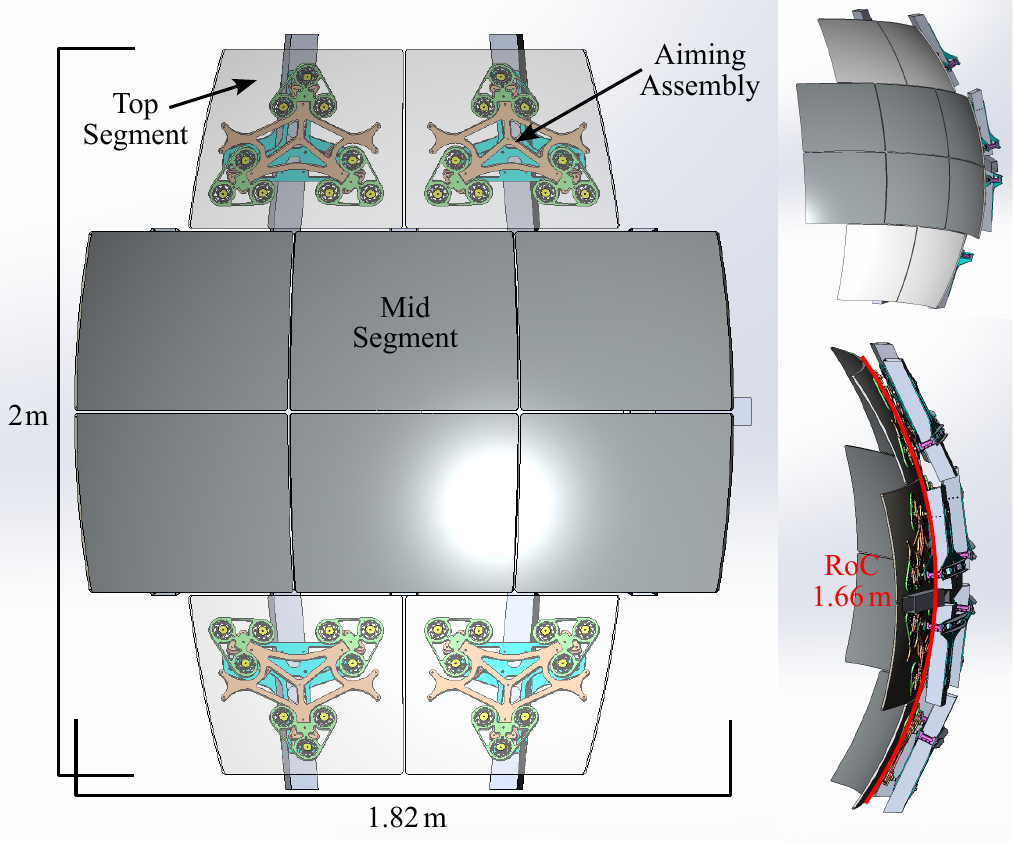}
        \caption{Drawing of the 10 segmented primary PBR mirror, including the aiming assembly on the backside.}
        \label{fig:Mirror}
    \end{minipage}%
\end{figure}

The telescope mechanical structure needs to support the optical design while preserving image quality across the full pointing range and the extreme environmental loads expected for a stratospheric balloon payload. The resulting modular design (see \autoref{fig:pbr-telescope}) comprises five primary components: the mirror assembly, the telescope (belt) frame, the aperture assembly, the camera shelf, and the dark‑box shell. The large square belt frame wraps vertically around the mirror just at the height of the telescope's center of mass; it bears the majority of the load, stiffens the optical train, and interfaces directly to the rotator. Two forward trusses on either side and a central truss carry the aperture assembly and camera shelf.
The aperture assembly secures the ACP, hosts the Infrared (IR) camera and X/$\gamma$ detector, and supports the top and bottom shutters. These shutters protect the telescope from direct sunlight and will be controlled by the operator from ground.  The mirror assembly holds the 10 segments in three‑point flexure mounts, shown in \autoref{fig:Mirror}. Nine Kovar pads are epoxied to each segment and the pads in turn attach to the aluminum structure through flexures and a Whipple‑tree distribution to equalize loads, accommodate differential thermal contraction and allow precise alignment of the each segment before launch. The camera shelf interfaces the FC and CC to the frame, permitting independent axial adjustment while minimizing obscuration.

To avoid stray light inside the telescope, we enclose the optics in a layered dark box: an outer reflective Mylar layer limits solar heating, a second layer of faraday fabric, a foam layer which provides structure and insulates against extreme temperatures, and an inner light‑absorbing surface suppresses stray light. The second main function of this enclosure is to provide EMI shielding for the RI. For this purpose, Titan RF Faraday\footnote{https://mosequipment.com/products/titanrf-faraday-fabric?variant=31133468983398} fabric will be added to each dark box panel and conductively connected to the telescope frame. To ensure strong suppression, joints between members and aluminum skins will be sealed with foam EMI gaskets and the cables passing through the darkbox will be filtered and jacketed. To complete the Faraday enclosure of the telescope, a copper mesh with a \unit[10]{mm} spacing of \unit[0.125]{mm} diameter wire is installed across the aperture and also conductively connected to the telescope structure. From these measures, we expect a suppression of more than \unit[45]{dB} across our frequency range on top of the EMI mitigation done for each electrical component individually.

The RI attaches to the telescope via a mast and frame mounted on the belt frame. Although the RI is not part of the optical assembly, this attachment keeps fields of view aligned for hybrid measurements and avoids obstructions by design.

\begin{figure}[!htb]
    \centering
    \includegraphics[height=5cm]{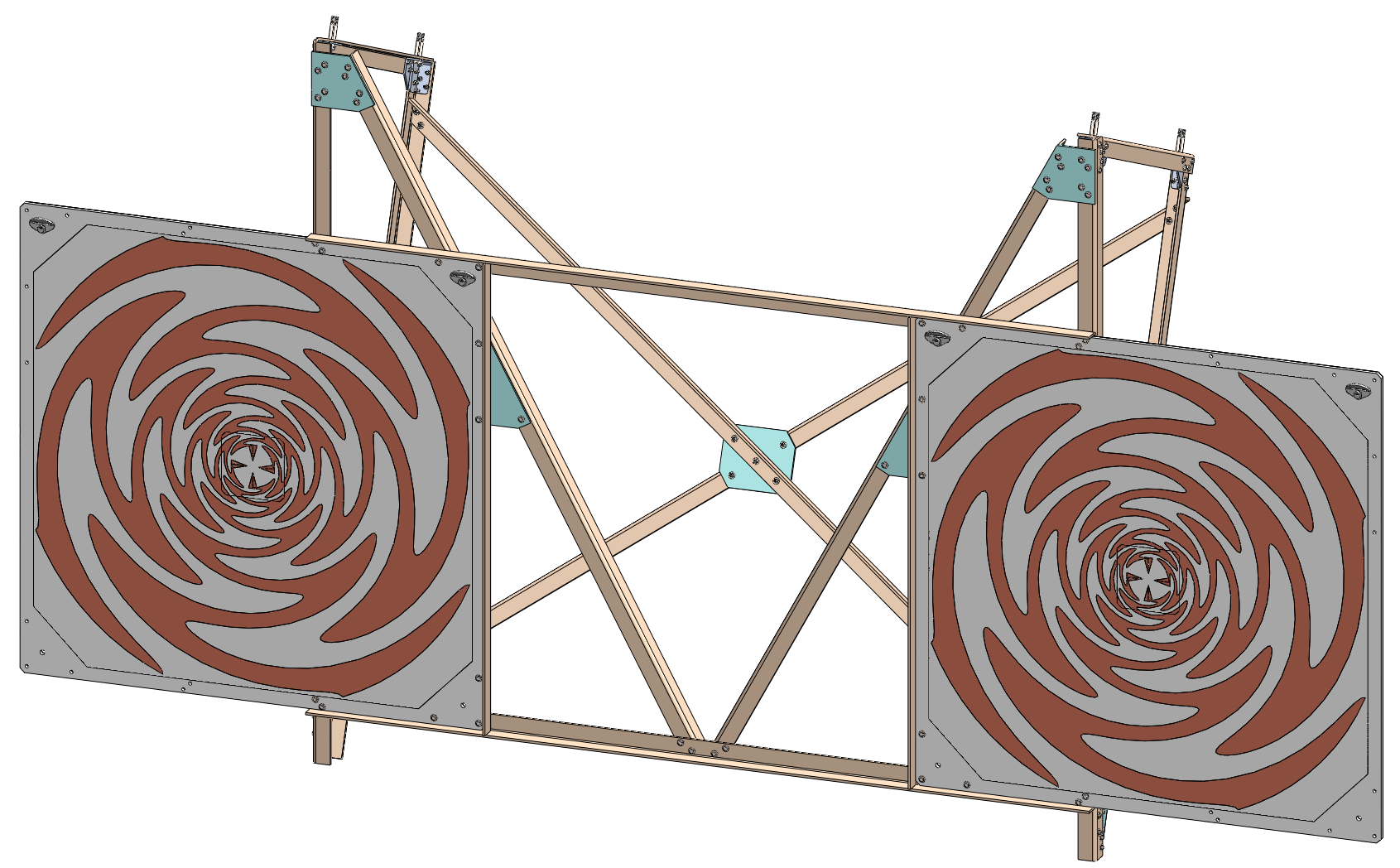}
    \caption{Radio frame and mast assembly.}
    \label{fig:radiomast}
\end{figure}

As shown in \autoref{fig:radiomast}, the mast positions the radio antenna 0.5\,m below the optical telescope, providing clearance from the gondola and preventing metal components from obstructing the antenna's $60^\circ\times60^\circ$ field of view or causing re‑emission interference. We designed the mast and frame to hold the antenna orientation within $1^\circ$ of the telescope optical axis across the full rotation range. The assembly uses minimal conductive material apart from fasteners and the radio electronics. We selected glass‑fiber‑reinforced vinyl (GFRV) for the mast because it is non‑conductive, radio‑transparent, and retains the required strength over the expected temperature range. The antenna support is a composite of FR4 blocking integrated into the antenna panels and GFRV bars that set a 1\,m inter‑antenna separation. Antenna elements use copper traces printed on FR4 and are bonded to an Aramid honeycomb for stiffness, producing a lightweight, rigid assembly with high gain.

\subsection{PBR Focal Surface}
\label{subsec:FS}
One major step towards a POEMMA like instrument is the novel realization of a hybrid focal surface. PBR's focal surface host roughly 10\% of the channels proposed for one of the POEMMA satellites. This focal surface consists of the Cherenkov camera, employing SiPMs and optimized for fast signals, and the Fluorescence Camera, utilizing Multi-Anode Photomultiplier Tubes optimized for the detection in UV on a $\mu$s timescale. The following paragraphs describe each camera in detail.

\begin{figure}[!ht]
    \centering
    \includegraphics[width=\linewidth]{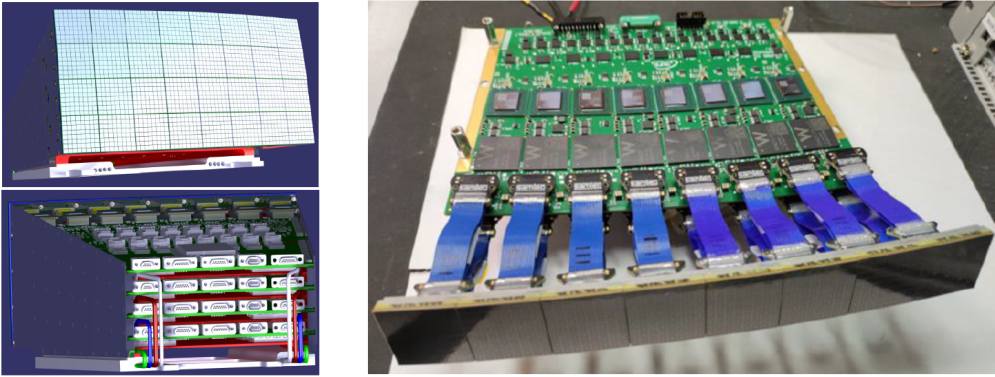}
    \caption{Left: CAD rendering (front view at the top, back view on the bottom) of the full Cherenkov Camera and electronics rack, illustrating the focal‑plane mounting and thermal enclosure. Right: schematic of the Cherenkov Camera elementary cell (CC‑PDM) showing the 8$\times$8 SiPM array, front‑end MIZAR ASICs, FPGA, and thermal‑control interfaces.} 
    \label{fig:CC}
\end{figure}

\paragraph{Cherenkov Camera (CC)} 
\label{subsec_para:CC}
The PBR Cherenkov Camera comprises four rows of eight 8$\times$8 SiPM arrays, each pixel measuring 3$\times$\unit[3]{mm\textsuperscript{2}}, for a total of 2048 pixels (\autoref{fig:CC}). The arrays mount on a structure that approximates the spherical focal surface required by the telescope's modified Schmidt optics (see \autoref{subsec:optics} for details of the optical system). The camera subtends a field of view of 12\textdegree$\times$6\textdegree, with each pixel covering 0.2\textdegree$\times$0.2\textdegree. The chosen sensor for the 8$\times$8 arrays is the Hamamatsu S13361‑3050, which is sensitive from 320 to \unit[900]{nm}.

The readout electronics is based on the next generation of application‑specific integrated circuit (ASIC) optimized for low power consumption and advanced functionality. The main candidate is the 64-channel CMOS MIZAR (Multi-channel Integrated Zone-sampling Analogue-memory based Readout) ASIC (developed by the INFN in Turin, Italy) \cite{DiSalvo:2023pbz}. Each MIZAR chip supports 64 channels and implements an integrated front‑end amplifier followed by 256 memory cells sampled at \unit[200]{MHz}. Each memory cell contains a storage capacitor and a single‑slope analog‑to‑digital converter with programmable resolution (7–12 bits); the memory can be partitioned in groups of 32, 64, or 256 cells to optimize dead time and buffering. The ASIC provides per‑channel thresholds, two programmable comparators, and on‑chip buffering so that digitization and readout occur only after trigger validation, minimizing data volume and dead time. Each MIZAR channel consumes on the order of \unit[10]{mW}. This low power consumption per channel is a critical step towards a space qualified camera.

The first results of the test of the MIZAR chip are encouraging, but the measurements to evaluate its full functionality are still in progress, so a possible alternative solution has been considered. The alternative solution is based on the Radioroc2 ASIC produced by Weeroc. 
While this option does not allow for wavelength digitization, it offers a lower data budget solution. The use of Radioroc2 ASIC enables channel-by-channel adjustment of SiPM High-Voltage, allowing for fine SiPM gain adjustment to correct for channel gain non-uniformity.  Additionally, this ASIC can trigger at levels as low as 1/3 photoelectron, offering dual-gain energy measurement capabilities, and maintains 1\% linearity in energy measurement up to 2000 photoelectron. The camera readout groups eight  ASICs on a single board (ASIC board). A second board, based on a Xilinx SoC/FPGA, manages the ASIC board. The eight SiPMs  array, an ASIC board, an FPGA board form the Cherenkov Camera Photodetection Module (CC-PDM). 
The CC‑PDM implements local trigger logic, controls memory partitioning, provides a temperature‑compensated bias supply for its SiPM array, and forwards trigger requests to a camera‑level trigger aggregator. A dedicated Trigger/Sync board (called clock board)
consolidates requests from multiple CC‑PDMs and issues the final trigger that initiates digitization and readout across the all CC-PDMs. Readout frames and housekeeping data stream from CC‑PDMs to the Data Processor over the instrument network; trigger counters and per‑event timing information (1PPS and GTU values) are appended to each event to preserve absolute and relative timing. 

The camera electronics reside in a thermally controlled rack designed to minimize electromagnetic interference and to maintain stable SiPM and front‑end performance under stratospheric conditions.

\paragraph{Fluorescence Camera (FC)}
\label{subsec_para:FC}

\begin{figure}
   \centering
   \includegraphics[width=\columnwidth]{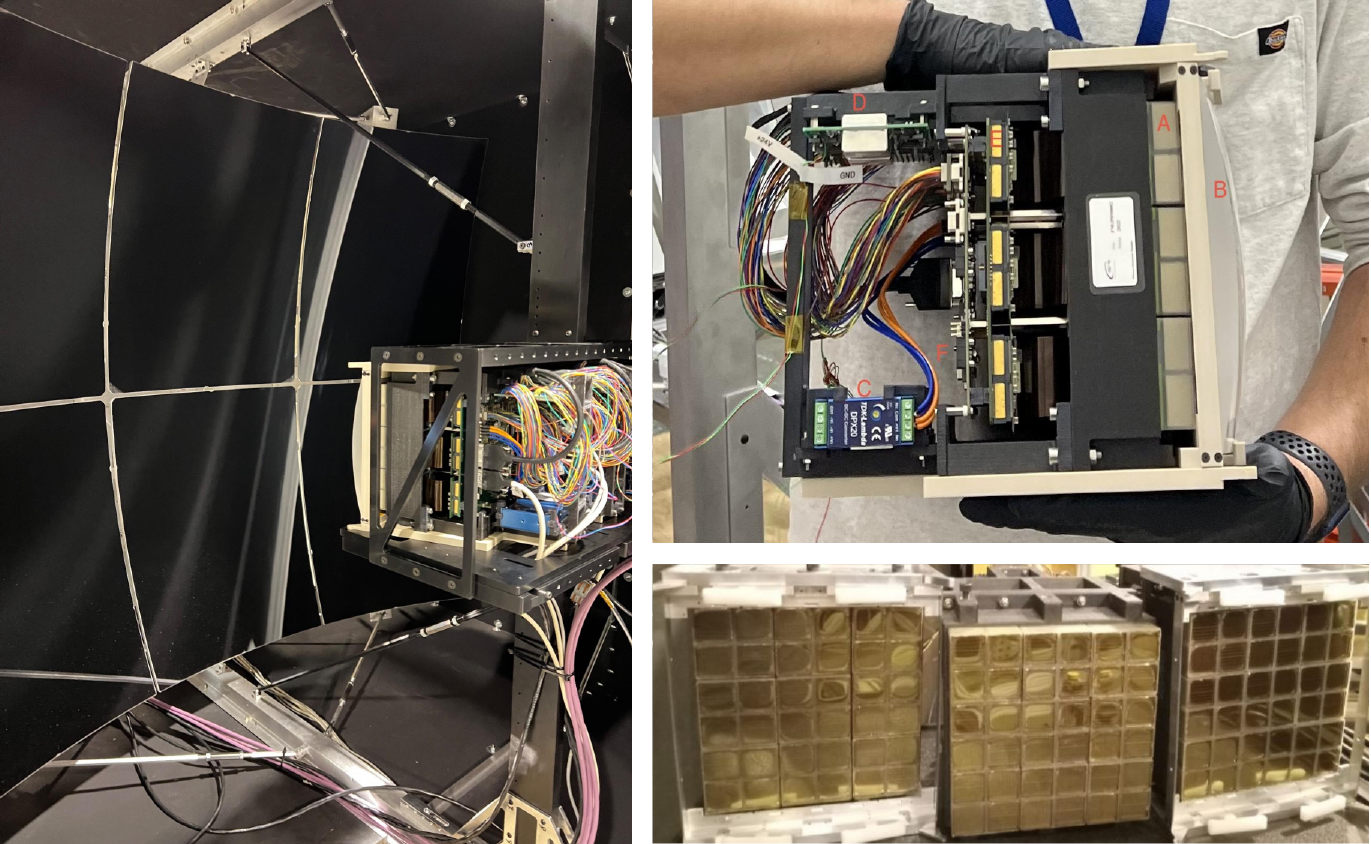}
   \caption{Photograph of the inside of the assembled EUSO-SPB2 Flourescence Telescope (left), the three PDMs in the laboratory without the lens assembly installed (bottom right), and one PDM with lens assembly installed held by a graduate student to show the scale (top right).\vspace{10mm}}
	 \label{fig:SPB2_PDM}
\end{figure}

The fluorescence camera (FC) comprises of four photo-detection modules (PDMs).
The four PDMs FC of PBR build on the EUSO-SPB2 FT \cite{SPB2_FT}.
An assembled EUSO-SPB2 PDM is shown in \autoref{fig:SPB2_PDM}.
A BG3 filter in front of each PDM constrains the wavelength range to between 290 and 430 nm, which reduces the background light. The PDM follows a modular design, built from nine elementary cells.
Each elementary cell shares a high-voltage power supply utilizing a Cockroft-Walton circuit \cite{Plebaniak:2017itg} and contains four 64-pixel multianode photomultiplier tubes (MAPMTs, Hamamatsu Photonics R11265-M64), to yield a 9216-pixel FC.
A SPACIROC3 ASIC \cite{Blin:2018tjp} performs single photoelectron counting on each pixel with a double pulse resolution of approximately 10~ns as well as charge integration on groups of 8 pixels to measure extremely bright signals. The integration time is 1 $\mu$s.
Each PDM is controlled by a Xilinx Zynq 7000 FPGA that handles triggering and data packaging. The proposed trigger scheme for PBR is based on the EUSO-SPB2 trigger  \cite{Filippatos:2022zfl}.
The trigger logic performed as expected in the flight of EUSO-SPB2, requiring only tuning the trigger parameters based on the instrument performance.
Prior to flight, the FC will be calibrated based on the methodology described in \cite{PARIZOT2025103112}.

This camera design has flown on EUSO-SPB1, EUSO-SPB2, and MiniEUSO \cite{Bacholle:2020emk}. Nevertheless, detecting a UHECR is still required; it was not achieved on these missions due to the short EUSO‑SPB1/2 flights and Mini‑EUSO's high threshold (well above $E_{thr}>10^{21}$~eV). This will be the first time the instrument measures an air shower signal and operates for a long duration in near space.
It will also be the first time that it is used as part of a hybrid focal surface. 

\subsection{Radio Instrument}
\label{subsec:RI}
The PBR radio subsystem adopts the Low Frequency (LF) drop‑down instrument developed for the PUEO as a template. The LF instrument is designed to be lightweight, energy efficient, and optimized for detection of radio emission from EAS \cite{PUEO:2020bnn}. PBR's RI is comprised of two dual‑polarized receiver channels, each consisting of a broadband sinuous antenna, a two‑stage front‑end RF signal‑conditioning chain, and supporting power and data acquisition electronics.

\begin{figure}[!h]
    \centering
    \begin{minipage}[t]{0.46\linewidth}
        \centering
        \includegraphics[width=\linewidth]{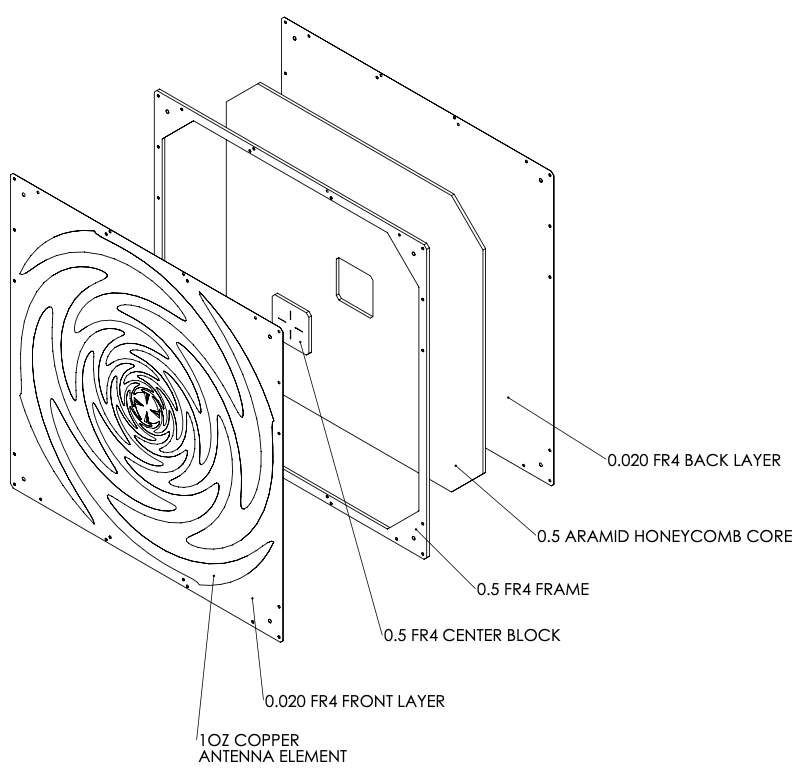}
        \caption{PBR antenna design}
        \label{fig:PBR_antenna}
    \end{minipage}%
    \hfill
    \begin{minipage}[t]{0.5\linewidth}
        \centering
        \includegraphics[width=\linewidth]{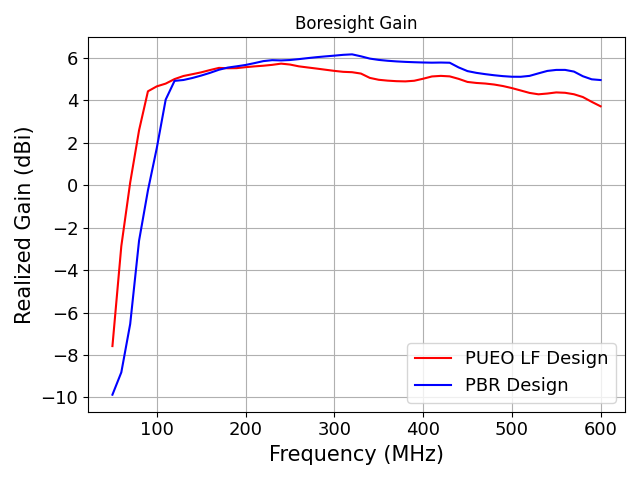}
        \caption{Simulated PBR antenna boresight gain, compared to PUEO LF design}
        \label{fig:PBR_antenna_gain}
    \end{minipage}
\end{figure} 

Figures \ref{fig:PBR_antenna} and \ref{fig:PBR_antenna_gain} show, respectively, the antenna concept and simulated boresight gain compared with the PUEO LF design. The PBR geometry differs from the PUEO design in order to increase gain and mechanical rigidity while meeting the payload's mass and envelope constraints. We will fabricate the antennas by printing copper traces on FR4 substrates with a circumscribed diameter of 52" (vendor limitation). Conductive vias interconnect the front and rear faces to simplify mounting and to support 4:1 impedance‑matching balun boards. An Aramid honeycomb support and the FR4 backing are effectively radio‑transparent and provide the required structural stiffness to preserve antenna pattern integrity during flight.

The antenna design provides approximately \unit[6]{dBi} broadband gain from \unit[60]{MHz} to \unit[600]{MHz} in both polarizations, with \unit[-3]{dB} beamwidths of roughly 
$\pm$30\textdegree in elevation and azimuth (see Fig.~\ref{fig:PBR_antenna_gain}). The two antennas are mounted beneath the optical telescope, co‑pointed with the Cherenkov Camera and separated by \unit[1]{m} to enable basic phasing and rudimentary beamforming (see \autoref{subsec:TelMech} for the mechanical details).

The front‑end electronics implement low‑noise amplification and band‑defining filtering. We place a radio amplifier board directly on each antenna rear face to minimize cable loss and follow with a RadioReceiver board and a RadioPower board for regulated supply. The chain delivers $\sim$\unit[60]{dB} of net gain across \unit[60–500]{MHz} while consuming less than \unit[1.3]{W} per channel at \unit[+5]{V}. We will calibrate the end‑to‑end gain and noise figure of each channel under representative thermal‑vacuum conditions to validate performance for balloon operations.

The data acquisition (DAQ) leverages an RFSoC (radio frequency system‑on‑chip) architecture similar to that used in PUEO. The RFSoC integrates high‑speed, multi‑gigasample analog‑to‑digital converters (up to \unit[5]{GSa/s}, 12–14 bit) with programmable logic and an embedded processing system. This architecture enables on‑chip antenna phasing, low overall power consumption ($\leq$\unit[25]{W} for the radio DAQ), and flexible interfacing to the flight computer. The RFSoC will implement a radio‑only trigger mode for daytime operation and will accept an external nanosecond‑scale trigger from the Cherenkov Camera via LVDS (main trigger mode). A software trigger for understanding measurement environments at any time will also be available. The RadioDAQ board provides eight coaxial inputs to receive the four differential RF channels (horizontal and vertical polarization for each antenna), LVDS lines for the external CC trigger and housekeeping, and an Ethernet interface to the flight computer for event and monitoring data. We limit the buffer latency between the CC trigger and the RadioDAQ to $\leq$\unit[130]{ns} and constrain per‑event radio data output to $\leq$\unit[20]{kB}. The design allows analog pre‑processing to be enabled or bypassed under RFSoC control via LVDS lines. We route power and RF signals using LMR240/195 coaxial cable selected for mechanical robustness and RF shielding.

\subsection{X- and Gamma‑ray Detector}
\label{subsec:XGamma}

\begin{figure}[!ht]
    \centering
\includegraphics[width=\textwidth]{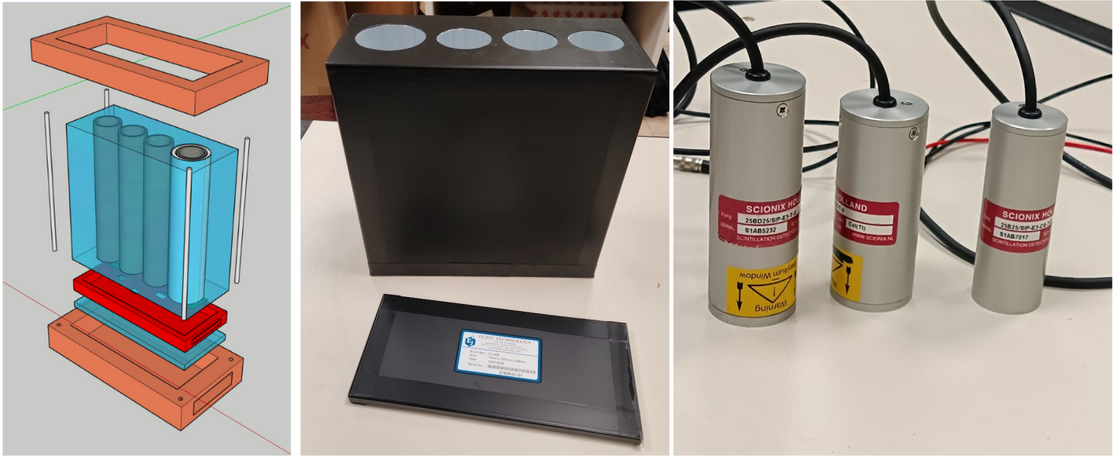}
    \caption{Left: CAD model of the X-$\gamma$ detector ($28\times16\times30$~cm$^3$). Cyan: anti-coincidence panels; orange: aluminum parts; red: PVC bases for lead collimators. Rightmost cylinder is the low-energy X channel (not covered by anti-coincidence to avoid absorbing low-energy X-rays). Aluminum enclosure omitted. Center: bottom view of the anti-coincidence assembly showing four collimator apertures; the smaller scintillator plate covers the openings. Right: three Scionix detectors used in the X-$\gamma$ detector. }
    \label{fig:XgammaDet}        
\end{figure}

The X- and Gamma-ray detector is located in the top left corner of the telescope (see \autoref{fig:pbr-telescope}). The system (see \autoref{fig:XgammaDet}) is composed of four Scionix modules that cover overlapping energy bands: a low‑energy X-channel ($<$\unit[20]{keV}), an intermediate X–$\gamma$-channel, above \unit[30]{keV}, and two $\gamma$-channels extending to several MeV, starting above \unit[50]{keV}. The modules share a common mechanical assembly and a field of view matched to the hybrid focal surfaces.
Each module uses a compact scintillating crystal (CsI(Tl) or NaI(Tl)) of roughly \unit[2.5]{cm} characteristic dimension, coupled to temperature‑compensated SiPMs. The detector modules are mounted inside a \unit[2]{mm} thick aluminum enclosure. The scintillator window is covered by a different entrance window material, appropriate for the corresponding channel: a thin beryllium window ($\sim$\unit[0.3]{mm}) for the low‑energy X channel and aluminum windows of different thicknesses, $\sim$\unit[0.1]{mm} for the intermediate channel and around \unit[1]{mm} for the high‑energy $\gamma$-channels). We define the $30^{\circ}$ FoV of the three $\gamma$-channels by enclosing them in lead collimators. To reduce the background for the X-channel, we limit its FoV to 16\textdegree. The collimator geometry and materials align the detector FoV with the optical instruments while suppressing environmental X‑ray background.

A plastic‑scintillator anti‑coincidence (EJ‑208) surrounds the whole detector assembly with the exception of the X‑channel entrance window, since it would stop X-ray photons in the energy range of interest (center panel of \autoref{fig:XgammaDet}). The anticoincidence scintillator employs SiPM readout (four channels total) to monitor charged‑particle flux; these signals are recorded in the data stream but do not act as an automatic veto because relativistic charged particles from HAHA showers can arrive at the same time with the photon signals. A dedicated, FPGA‑based computer controls acquisition, 
and temporary storage prior to transmission to the main payload computer.
Upon the detection of a signal over threshold, the values are read out and stored. In parallel, the system monitors the rate of both the four main channels and the four anticoincidence SiPMs. This rate will be used to measure particle fluxes at different geomagnetic latitudes and altitudes. Such measurements are still relatively scarce in the literature, particularly at the typical float altitudes of super-pressure balloon missions.
In addition, the system can receive trigger notification from the other subsystems, to enable an offline analysis of synchronous events.

\begin{figure}
   \centering
   \includegraphics[width=.66\columnwidth]{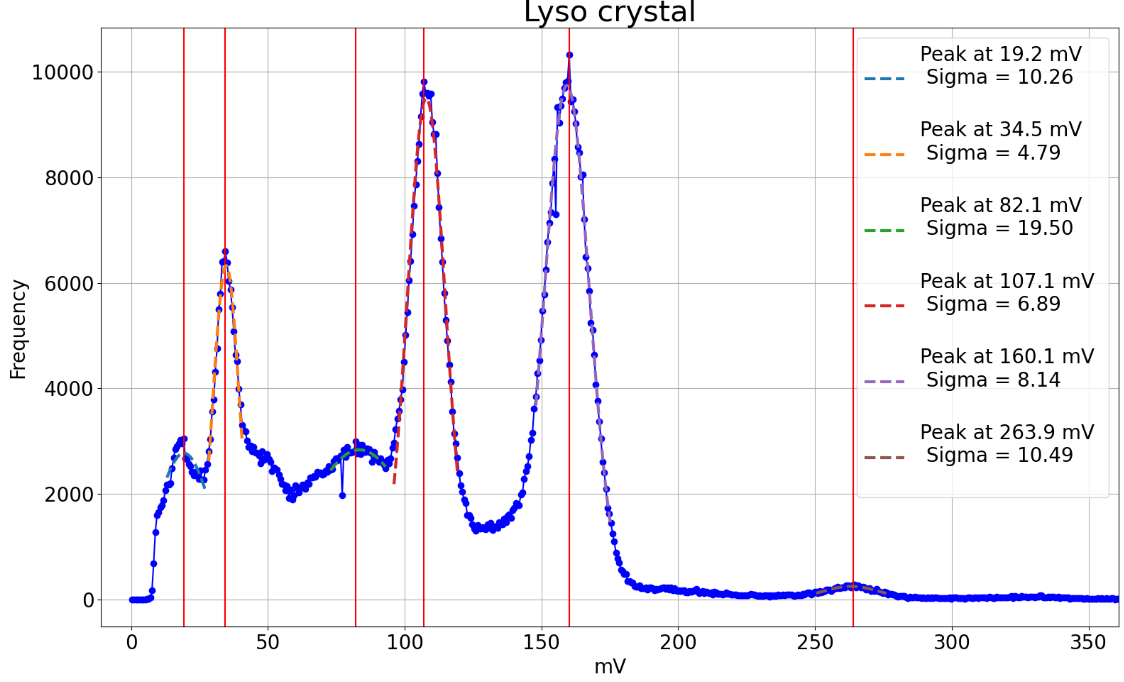}
   \includegraphics[width=.33\columnwidth]{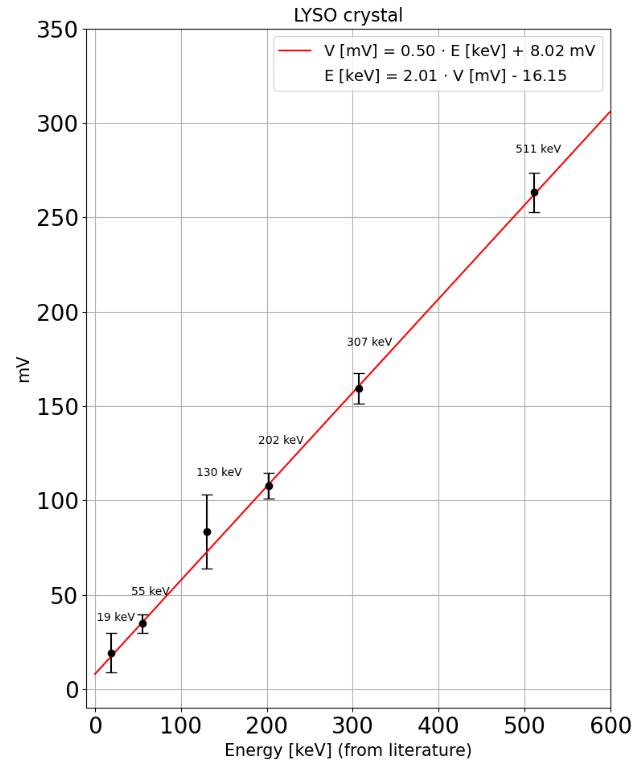}
    \caption{\textbf{Left:} Energy spectrum measured with a lutetium-containing LYSO (lutetium–yttrium oxyorthosilicate) crystal using the X-channel. The distribution shows the event counts as a function of the signal amplitude (in mV), with clearly identifiable peaks associated with known radioactive emissions. \textbf{Right:} Linearity of the detector response. The plot shows the correlation between the measured signal amplitude and the corresponding photon energies, as reported in the literature for the identified spectral lines. The red line represents a linear fit to the data.}
	 \label{fig:LYSO}
\end{figure}

Laboratory tests on the engineering model, with a first-generation electronics demonstrate an effective energy threshold below 20~keV. (Fig.\ref{fig:LYSO}, left) and a linear energy response in the range 20~keV to 500 keV for the low-energy X-channel. The linearity of the detector response was evaluated by fitting the measured peak positions as a function of the corresponding photon energies. A weighted linear fit yields a slope of $(2.01 \pm 0.09)$~keV/mV, with a reduced $\chi^2$ close to unity. The deviations from linearity are found to be below a few percent over the explored energy range.
Spectra from several other radioactive sources have been measured with similar results.

\subsection{Data processing system} 
\label{subsec:DP}

\begin{figure}[!h]
    \centering
    \includegraphics[width=\linewidth]{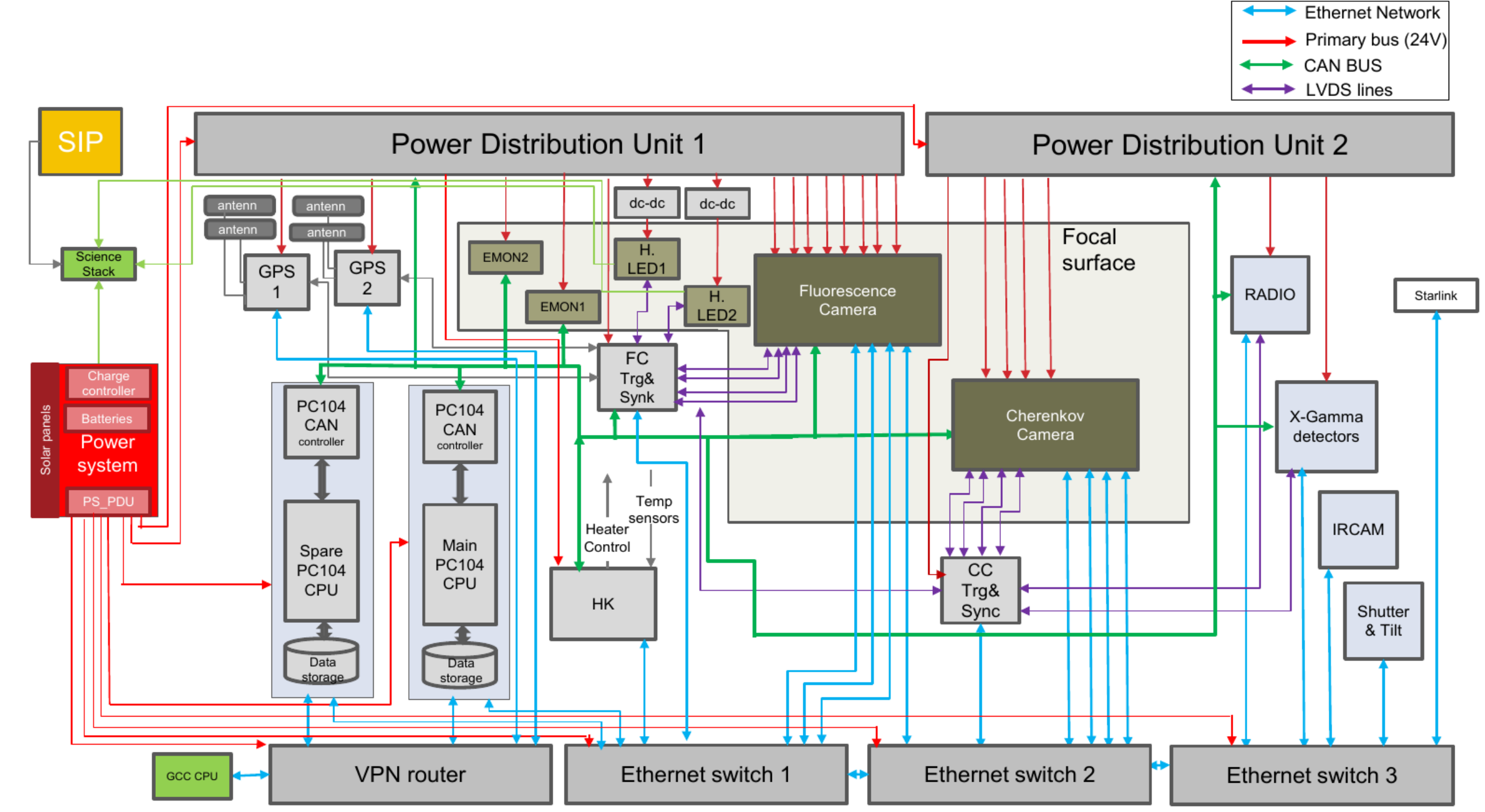}
    \caption{Block diagram of the PBR Data Processor showing interfaces to all the sub-systems and telemetry (SIP), see \autoref{subsec_para:telemetry}; the diagram illustrates our approach to real-time data handling, and synchronization across all instruments.}
    \label{fig:PBR-DP}
\end{figure}

The PBR Data Processor (DP) synchronizes data acquisition across all instruments, manages trigger decisions, and handles storage and telemetry of science data. It builds on the EUSO‑SPB2 DP architecture and incorporates upgrades to support the hybrid focal surface, the radio instrument, and the gamma/X‑ray detector. The DP electronics reside in a customized Eurocard chassis with a cooling plate; commercial off‑the‑shelf components were selected and qualified for extended stratospheric operation to maximize reliability. The block-diagram in \autoref{fig:PBR-DP} illustrates the full system layout.

The primary DP hardware comprises two PCI/104 single‑board computers configured as hot/cold redundancy connected to five SATA disks providing 4\,TB of storage. Synchronization and trigger functions are performed by trigger/synchronization boards based on a Xilinx XC7Z SoC. One is dedicated to the FC only, while the second one handles the remaining sub-systems, it distributes the 1‑pulse‑per‑second (1PPS) GPS timing and the global GTU clock, interfaces to the GPS receivers, and manages local and global triggers. The DP also hosts Ethernet switching and a solid‑state power controller for subsystem power distribution.

The DP enforces sub‑microsecond alignment among detectors. The master CC Trg\&Sync board distributes trigger signal, 1PPS and GTU counters, and busy signals; each global trigger records the 1PPS and GTU values together with level‑1 and global trigger counters, which are appended to event packets to preserve absolute and relative timing. Subsystems include live and dead counters with each event so that the summed live+dead time equals the run duration.

We implement a hierarchical trigger architecture to accommodate diverse, asynchronous signals. Each subsystem generates local triggers independently. The DP time‑stamps trigger requests, applies coincidence windows and acceptance logic, and, on validation, initiates readout, assembles a unified event packet with metadata, and applies cross‑instrument timing offsets and calibration constants.

The DP provides the interface between detectors and telemetry and supports distributed operations. Control software configures and monitors subsystems and exposes command/telemetry interfaces; data‑handling software performs acquisition, prioritized buffering for downlink, and health validation.

The onboard software is modular in design and performs operations in asynchronous mode. Based on a \textit{client-server} architecture, the software main components are manager applications which handle the exchange of commands and data in a fully configurable fashion. They handle Ethernet, CAN-Bus and Serial links, using a common framework that is fully configurable enabling real-time definition of command sequences that can be executed once or accordingly to user defined time sequences. All these applications share the same configuration module that allows an operator to add also data consumers sub-processes that can perform local or remote logs, store data in files or databases and perform data sanity checks. Thanks to its modular design, it can be configured to be used in every phase of the detector integration and commissioning. 

The onboard software is complemented by a ground sector that implements the data collection and detector monitor, along with the control dashboard to enable the mission operation by collaborators from different geographic locations. 

\subsection{PBR Infrared Camera}
\label{subsec:IR}

\begin{figure}[!ht]
    \centering
    \includegraphics[width=.75\linewidth]{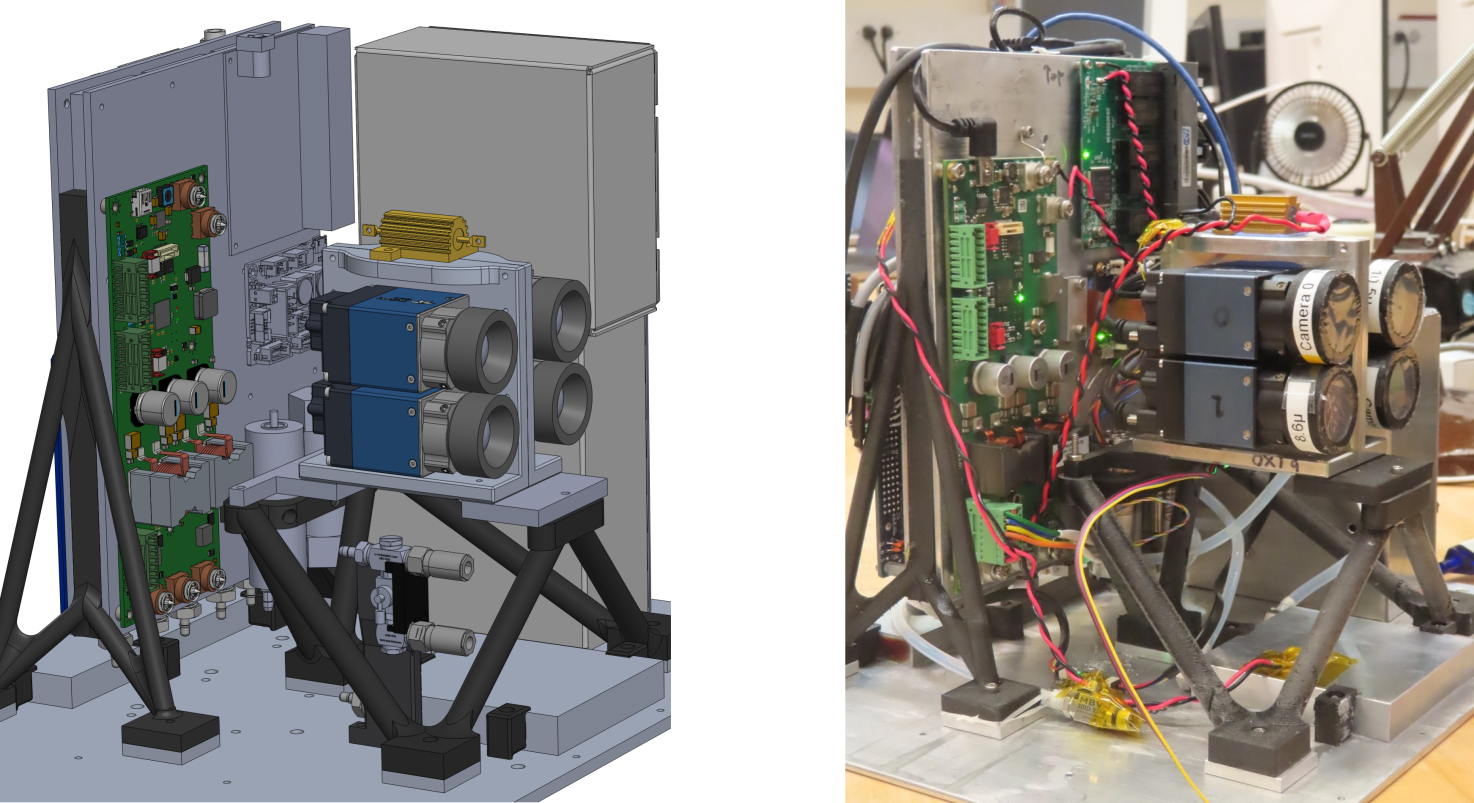}
    \caption{Left: Design drawing of the IR camera system for PBR. Right: Photograph of the system before T/Vac testing; the enclosure is not shown here.}
    \label{fig:UCIRC3}
\end{figure}

Quantifying cloud coverage within the PBR telescope field of view is essential for correct interpretation of the optical measurements. We designed and manufactured an updated infrared (IR) camera system (see \autoref{fig:UCIRC3}) based on the concept flown on EUSO‑SPB2 \cite{Diesing:2023wcq} and optimized it for PBR operational requirements. The system, located in the top right corner of the telescope enclosure (see \autoref{fig:pbr-telescope}), comprises four Teledyne DALSA Calibir GXF cameras. Three cameras employ narrowband filters centered at 8.6, 10.5 and \unit[12.5]{$\mu$m}, respectively, while the fourth records broadband radiance across the \unit[8–14]{$\mu$m} range. Combining narrowband and broadband measurements improves cloud‑top temperature retrievals and supports cloud‑height estimation from differing viewing angles.

We redesigned the electronics enclosure to minimize electromagnetic interference and to permit installation inside the telescope so that the IR field of view co‑aligns with the one of the hybrid optical focal surface. We integrated an active cooling system to stabilize camera temperatures over a broad range of conditions to improve the camera calibration. During operations the system will acquire one frame every two minutes; these frames will support pixel‑level determination of cloud temperature and optical properties and provide estimates of cloud‑top height.

PBR Infrared Camera system is able to determine the cloud top height within \unit[500]{m} from the measured cloud top temperature when pointed in the nadir direction. The instrument is calibrated to absolute temperature in a thermo-vacuum test chamber by taking images of a black thermally controlled target under conditions spanning those expected in flight. A calibration model over the range of camera temperatures and range of atmospheric pressures encountered at float altitude had been fit to the calibration data. The model results in a maximum \unit[3]{$^\circ$C} calibration error for each pixel of the cameras. The atmosphere is modeled and numerically integrated through the atmosphere by the publicly available coupled ocean–atmosphere radiative transfer (COART) program described in \cite{ji02200w}. In an earlier flight, a similar system with only the 10.5 and 12.5 $\mu$m wavelength cameras permitted us to determine the location and the difference between high and lower clouds limited by the calibration accuracy of the previous version. The current improved calibration along with the added narrow band at 8.6 microns as well as an unfiltered camera accepting the whole range from 8 to \unit[12.5]{$\mu$m} wavelength is modeled to permit the \unit[500]{m} uncertainty in the cloud tops.

On the other hand with the IR camera system mounted on the telescope and pointed aligned with the main camera, we have no experience in the case where the camera is pointed at the limb or some other nadir angle. The COART model permits the numerical evaluation of the atmosphere other angles but we have not assessed the uncertainty in cloud distance or height during times when we are pointed at the earth limb.

\subsection{Auxiliary Systems}
\label{subsec:auxsystems}

\paragraph{Telemetry}
\label{subsec_para:telemetry}
The payload will use a low-rate Iridium link (255 bytes/minute) for primary command and state monitoring via the CSBF SIP. We will use the NASA Tracking and Data Relay Satellite System (TDRSS) with bursts up to 130 kbps (peak rates depend on orbital geometry and scheduled contacts) mainly for backup data transfers.

The main science downlink will rely on two maritime Starlink units providing redundancy for this link. Starlink first flew on NASA stratospheric balloon missions in 2023 (SuperBIT \cite{Gill_2024} and EUSO‑SPB2) and is currently under evaluation by NASA's balloon program office. Each unit supports sustained data rates on the order of hundreds of megabits per second, which should enable retrieval of the majority of science data even in the event of failed payload recovery. We will exercise the Starlink units extensively in preflight tests to characterize long‑term performance under balloon environmental conditions.

To minimize interference and simplify operations, the Starlink will be isolated from the TDRSS and Iridium links on the network and the power side. This separation minimizes the risk of interference of this new link with flight proven systems.

\paragraph{Gondola Control Computer}
\label{subsec_para:GCC}
The Gondola Control Computer (GCC) resides in a sealed, \unit[1]{atm} enclosure to improve thermal control. Its design derives from the GCC flown on EUSO‑SPB2, which performed as expected during that flight. The PBR GCC includes a larger hard drive to archive scientific data onboard and to buffer data between download opportunities. We upgraded the GCC network switch to a \unit[1]{Gbps} unmanaged unit to increase onboard communication bandwidth. The GCC monitors and controls the power system and handles payload telemetry (see telemetry section). The GCC software has been enhanced to support larger file sizes, to transfer data via Transmission Control Protocol (TCP) in addition to User Datagram Protocol (UDP), to utilize the SIP high‑rate science port, and to operate multiple, rate‑limited telemetry pathways concurrently.

\paragraph{Power system}
\label{subsec_para:PS}
The power system follows the design used aboard EUSO‑SPB2 with minor adaptations to meet new requirements. The system uses Li‑Ion batteries recharged by a solar array that can point to the Sun using the NASA rotator (see \autoref{subsec:gondola_pointing}), requiring only one science solar array for nighttime operation. The current design requires 15 custom-made \unit[69]{cm}$\times$\unit[79]{cm} solar panels from SBM Solar, Inc.; each panel produces \unit[110]{W} at float. 
Three panels wired in series form one array. Three such arrays are wired in parallel to a single charge controller, and two arrays are wired in parallel to a second charge controller. The charge‑controller outputs feed the \unit[25.6]{V} direct‑current battery pack, which comprises seven \unit[100]{Ah} Li‑Ion batteries from Global Technology Systems, Inc., connected in parallel. Given Li‑Ion performance at low temperatures, we sized the system to supply \unit[900]{W} for 15 hours of darkness while maintaining a state of charge above 20\%.

\paragraph{Health Light‑Emitting Diodes (HLED)}
\label{subsec_para:HLED}
PBR carries a redundant set of HLED units to monitor instrument response throughout the mission. Each unit emits a calibrated flash pattern; the outputs are offset by \unit[16]{s} and synchronized to the one‑pulse‑per‑second signal from the payload GPS. Each unit drives two LEDs that operate on different timescales: one produces microsecond‑scale pulses for the FC and the other produces nanosecond‑scale pulses for the CC. The HLED system includes an integrated temperature sensor whose readings support offline, temperature‑dependent amplitude corrections.

\begin{figure}[!ht]
    \centering
    \includegraphics[width=.75\linewidth]{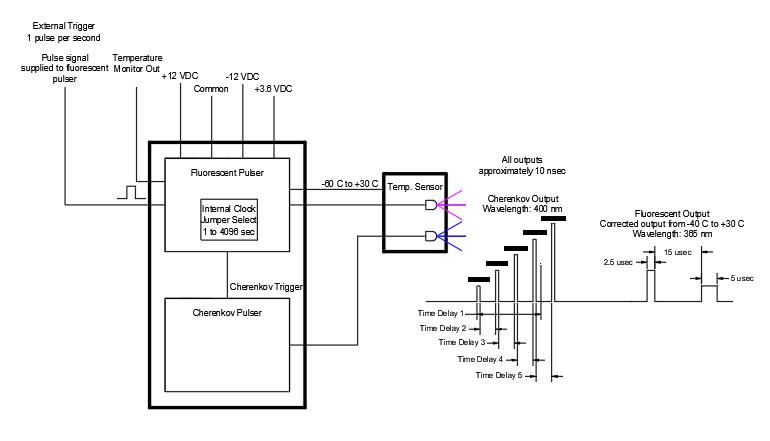}
    \caption{Layout and timing of a Health Light‑Emitting Diode (HLED) unit. Left: schematic of inputs and outputs, including power, GPS 1‑pulse‑per‑second (1PPS) synchronization, temperature sensor, dual LED drivers, and the LEDs themselves. On the right the GPS‑synchronized pulse pattern produced by the unit is shown.}
    \label{fig:placeholder}
\end{figure}

The LED used to monitor the FC emits at \unit[365]{nm} and produces pulses of \unit[2.5]{$\mu$s} and \unit[5]{$\mu$s} duration separated by \unit[15]{$\mu$s}; each pulse delivers approximately 500 photons at the detector. Because of the short pulse durations, the CC monitoring LED (\unit[400]{nm}) emits five \unit[10]{ns} pulses with an adjustable inter‑pulse spacing; the pulses have nominal amplitudes of 10, 30, 100, 300, and 500 photons per pulse. We record detector temperature during operation and apply a temperature‑dependent calibration correction on the ground.

The HLED subsystem is independently powered and operates autonomously (it is not controlled directly by the Data Processor). It accepts the cameras' internal trigger for normal operation and also provides an external trigger output to the Data Processor for event tagging and forced readout when required.

\paragraph{EMON}
\label{subsec_para:EMON}
Telescope Monitor (EMON) is a compact auxiliary monitoring device designed to measure low‑level light conditions within the main telescope environment. The instrument utilizes a Hamamatsu Si photodiode (S1227-1010BQ) coupled with particularly designed and developed onboard signal‑conditioning electronics to convert incident light into calibrated power measurements. Acquired data are periodically transmitted to the host system through a CAN communication interface. Its small size (\unit[55]{mm}$\times$\unit[40]{mm}$\times$\unit[30]{mm}), low power consumption ($<$\unit[1]{W}), high sensitivity (50-\unit[4000]{pW/cm$^2$}), and an electromagnetically shielded enclosure make it a suitable auxiliary monitoring device for the mission.
\section{Expected performance of the PBR payload}
\label{sec:expPerformance}

\subsection{Expectation for UHECR Observations}
\label{subsec:expPerformance_UHECR}

The rate for PBR, estimated by detailed simulations, is expected to be significantly higher (more than twice) than for previous missions due to the 25\% increase in the FC field of view (FoV) and increase in aperture size reducing the energy threshold of the FC. Reconstruction techniques have been developed in the JEM-EUSO collaboration for previous experiments and the same framework and tools are applicable to PBR.

\begin{figure}[h!]
\includegraphics[width=0.5\textwidth]{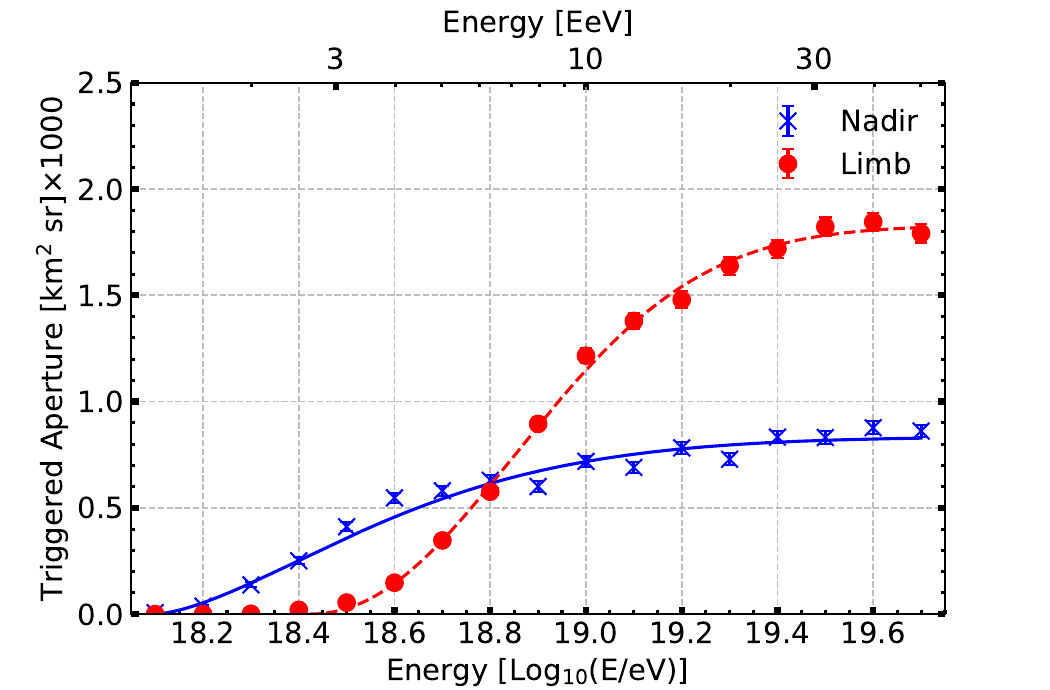}\includegraphics[width=0.5\textwidth]{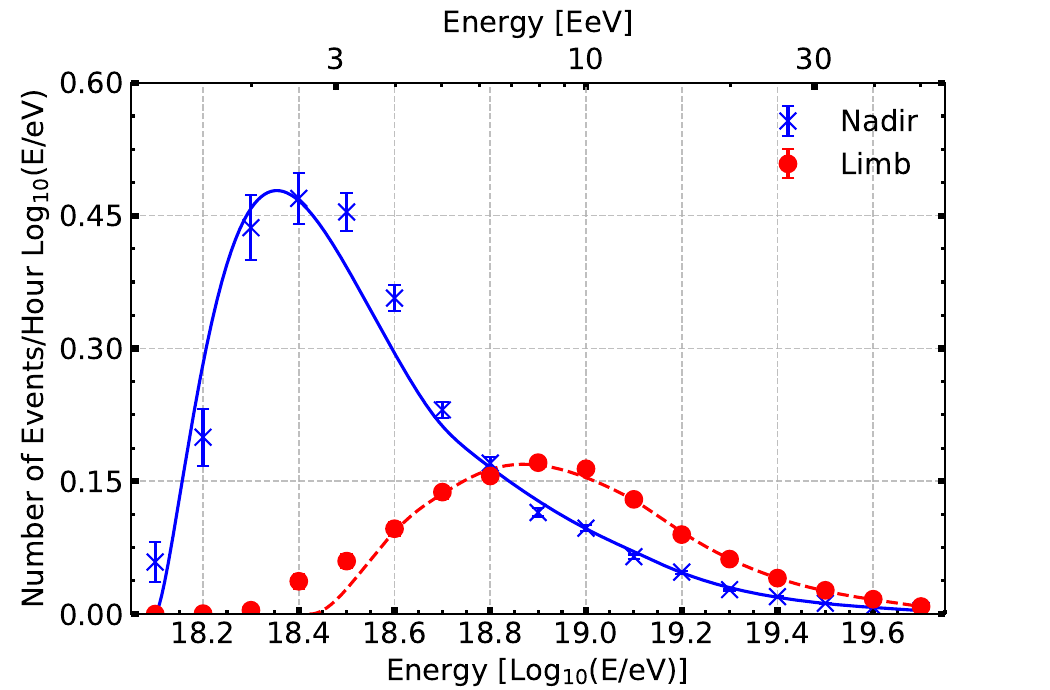}
\caption{Simulated triggered aperture (left) and event rate (right) for the FC based on Monte Carlo simulations. Showers are thrown in discrete energy bins shown by the markers with 100,000 showers per energy bin. The triggered aperture is converted to an expected event rate via the energy spectrum reported by the Pierre Auger Observatory \cite{PhysRevD.102.062005}.}
\label{fig:fc_sims}
\end{figure}

The simulated expected performance of the FC is shown in \autoref{fig:fc_sims}. The nadir case has the telescope's optical axis pointing straight downwards, while the limb case has the optical axis pointed 25$^\circ$ below horizontal which is a common configuration for the telescope to be in while the CC is searching for astrophysical neutrinos. These expectations are based on a large-scale Monte Carlo simulation using 2 million thrown showers for each observing geometry. These simulations are carried out in the EUSO-\Offline{} framework \cite{JEM-EUSO:2023fyg} with the methodology described in detail in \cite{SPB2_FT}. 

The triggered aperture is highly energy dependent since higher energy showers can be seen from further away from the detector. The energy threshold increases during limb pointing, as the showers which are in the FoV are generally further away from the detector. However, a significantly larger volume of atmosphere is observed resulting in a larger aperture at the highest energies. While observations are expected at higher telescope elevation angles, simulations reveal that the aperture does not increase above the limb case (optical axis pointed 25$^\circ$ below horizontal), even at the highest energy simulated (50 EeV). Integrating the curves of the right panel of \autoref{fig:fc_sims} the total number of expected triggered events per hour of observation is $0.28\pm0.02$ when the telescope is nadir and $0.12\pm0.01$ when the telescope is pointed at the limb.\\
A successful EAS measurement will increase the FC TRL to TRL6, the level required by space agency for their missions and hence an important step towards POEMMA.


\subsection{Expectation for High Altitude Horizontal Air-showers}
\label{subsec:expPerformance_HAHA}

\begin{figure}[!htb]%
    \vspace{-2mm}
    \centering
    \subfloat {{\includegraphics[width=0.5 \linewidth]{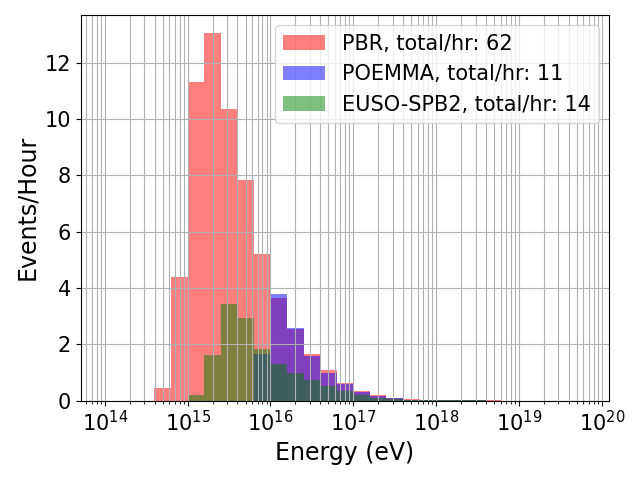} }}%
    \subfloat {{\includegraphics[width=0.5 \linewidth]{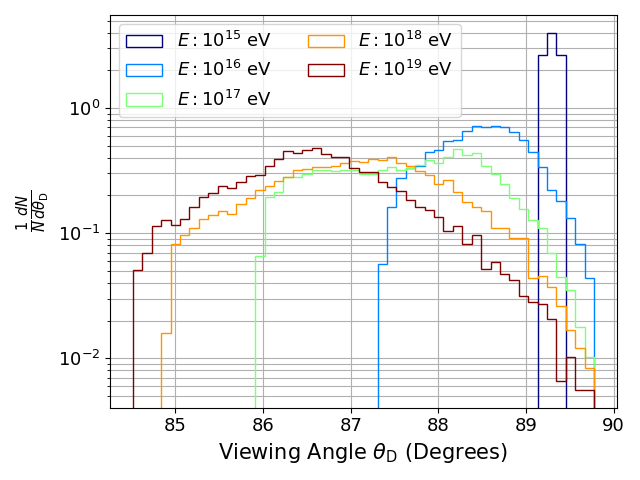} }}%
    \caption{Left: Total expected number of events for PBR, POEMMA, and EUSO-SPB2 per fifth-decade bins in energy. PBR will significantly increase the number of observed events as compared to EUSO-SPB2. Right: Normalized angular distribution of accepted PBR HAHA events for different primary cosmic ray energies.}%
    \label{fig:pbr_events}%
    \vspace{-2mm}
\end{figure}

Using the \texttt{EASCherSim}~\cite{Cummings:2020ycz, Cummings:2021bhg} optical Cherenkov simulation code, we have generated the properties of the Cherenkov emission for HAHA type events observed at 33~km altitude with viewing angles $\theta_{d}$ ranging from $84.2^{\circ}$ (Earth's horizon) to $92.2^{\circ}$. \autoref{fig:pbr_events} shows the estimated HAHA event rate for the configuration of PBR's CC as well as the normalized angular distribution of accepted events as a function of energy. The simulated event rate for EUSO-SPB2 and POEMMA are also shown to respectively demonstrate the improvement in the CC design and the shift to higher energies for a space-based mission. The CC of PBR is expected to observe roughly one HAHA event per minute of live time, with exposure down to $\sim 500$~TeV, and a maximum sensitivity around 3~PeV. The right panel of \autoref{fig:pbr_events} demonstrates that, due to atmospheric attenuation, HAHA events induced by higher energy primaries are expected to arrive with a wide range of angles while the majority of events are expected in the upper FoV of the camera. This is consistent with the HAHA candidate events detected by the short-duration flight of EUSO-SPB2, which arrived primarily in the bottom pixels of the camera after reflection through the optics \cite{Eser:2023lck, Cummings:2023ypo}. Thus, the majority of detected events are expected to probe early to middle ($s = 1$) shower development, with the instrument inside active shower development for event trajectories of $\theta_{d} > 87^{\circ}$. The properties of the Cherenkov photons arriving from a HAHA event mimic those generated by an upgoing, neutrino-sourced EAS. In this way, HAHAs present a guaranteed proxy signal by which to evaluate and refine the detection technique toward spaced-based observation of neutrinos for the POEMMA mission.

Recent work has shown that the energy threshold of a radio instrument to EAS can be lowered to a few \unit[10]{PeV} by machine learning techniques trained on simulations of the EAS radio pulses and background recorded by forced triggers to a few \unit[10]{PeV} \cite{Rehman:2023jme}.
As HAHA radio emission has not yet been well studied, it is difficult to predict the exact threshold for PBR in advance, and investigating the radio emission from HAHAs is a science goal in itself.
Due to the external trigger by the Cherenkov instrument, we will record air-shower radio events of interest even if they are below the analysis threshold of existing techniques.

Once HAHA radio emission is well understood, e.g., if the predictions of simulations can be confirmed experimentally, we can apply that knowledge to a second-level analysis of radio-Cherenkov hybrid events.
Coincident measurement of radio emission and optical Cherenkov emission from a single HAHA event can then provide a measure of both the angle of the Cherenkov cone and the angle of the shower trajectory with respect to the geomagnetic field through the measurement of the frequency spectra and polarity of the radio emission, respectively \cite{Schoorlemmer:2015afa}. These measurements will help to constrain the inference of the cosmic-ray mass composition for HAHA event trajectories where $X_{\mathrm{max}}$ is observable for $\theta_{d}<88^{\circ}$. Coincident measurements in optical Cherenkov and radio also effectively eliminate single-channel anthropogenic noise sources and can either rule out or confirm the steeply upgoing and near-horizon anomalous events observed by ANITA \cite{ANITA:2016vrp, ANITA:2018sgj, ANITA:2020gmv}.

The simultaneous observation of high-energy photons  produced during the development of HAHAs with the Cherenkov camera and the Radio Instrument allows PBR to reconstruct the different stages of the shower development using independent and complementary techniques, providing sensitivity to complementary regions of the shower phase space.

The X/$\gamma$ detector enables measurements that were not previously accessible, offering a unique probe of the initial phase of air‑shower evolution. However, realizing this potential is experimentally challenging: the observable photon flux depends on several factors, the very early stages of shower development when high‑energy electrons are produced, the absorption of photons during propagation, and the lateral spreading of the signal over large areas, each subject to substantial, model‑dependent uncertainty. At the same time, the detector will  monitor  the environmental $\gamma$-ray emission of cosmic and atmospheric origin. Based on previous measurements~\cite{Cumani2019, Winkler2003INTEGRAL}, a typical background rate of a few Hz per channel is expected.

Due to uncertainties in photon production, atmospheric absorption, shower-to-shower fluctuations, and limited knowledge of the ambient $\gamma$-ray background at float altitude, the actual detection sensitivity cannot be established a priori. Nonetheless, the X/$gamma$ detector remains the best available option to probe these poorly understood aspects of early shower development in a rarefied atmosphere.

\subsection{Expectation for Neutrino search from Targets of Opportunity}
\label{subsec:expPerformance_ToO}


An end-to-end software package, $\nu$SpaceSim, has been developed for determining neutrino diffuse flux and ToO fluence sensitivities \cite{Krizmanic:2023pwf,Reno:2025tpw}. Modeling of the EUSO-SPB2 CC is in progress using the EUSO-\Offline{} software \cite{JEM-EUSO:2023fyg}, and will be adapted for the PBR CC. A full implementation of a simulated PBR instrument is near maturity as a Geant4 package together with other physics simulation modules, working in concert within the EUSO-\Offline{} package. In addition, a modular and open-source software, NuTS\footnote{https://pypi.org/project/too-nuts/} \cite{Heibges:2025llt, Heibges:2023yhn, Wistrand:2023mpb, Posligua:2023cdm}, has been developed to monitor alerts and develop a catalog of potential very-high energy neutrino sources. By accounting for the balloon trajectory and observing conditions, NuTS determines lists of observable sources and produces optimized nightly schedules for ToO observations. The schedules prioritize likely sources of VHE neutrinos \cite{Guepin:2022qpl,Venters:2019xwi}, the length of the observation time, and optionally, whether or not the transient source has already been observed during the mission. Observation schedules are produced every night before the beginning of the observation window and can be easily modified if new high priority alerts happen during the night.

\begin{figure}[!htb]%
    \vspace{-2mm}
    \centering
    \subfloat {{\includegraphics[width=0.5 \linewidth, trim = 9.4cm 30mm 9.5cm 30mm, clip]{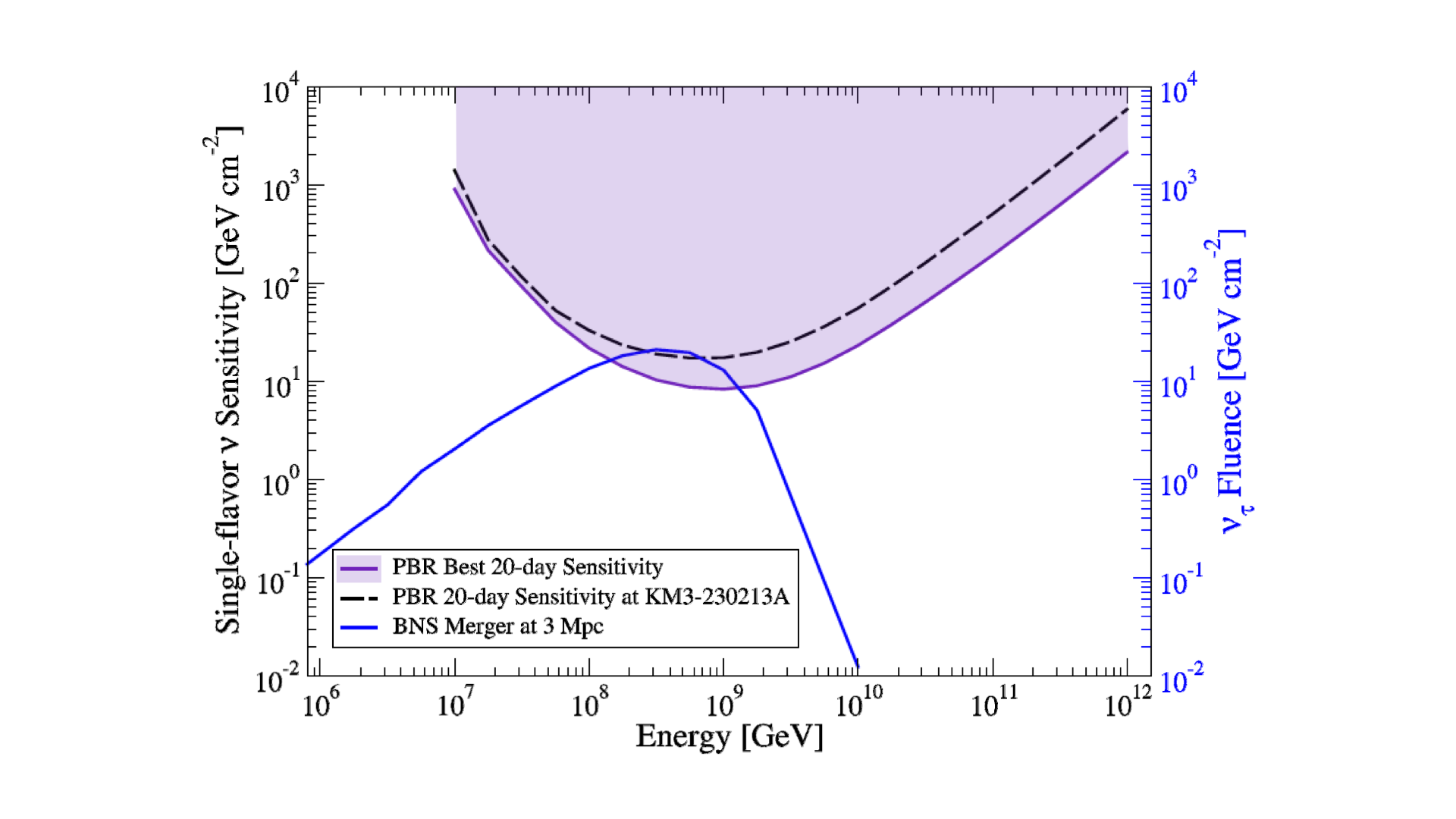} }}%
    \subfloat {{\includegraphics[width=0.5 \linewidth, trim = 9.4cm 37mm 9.5cm 33mm, clip]{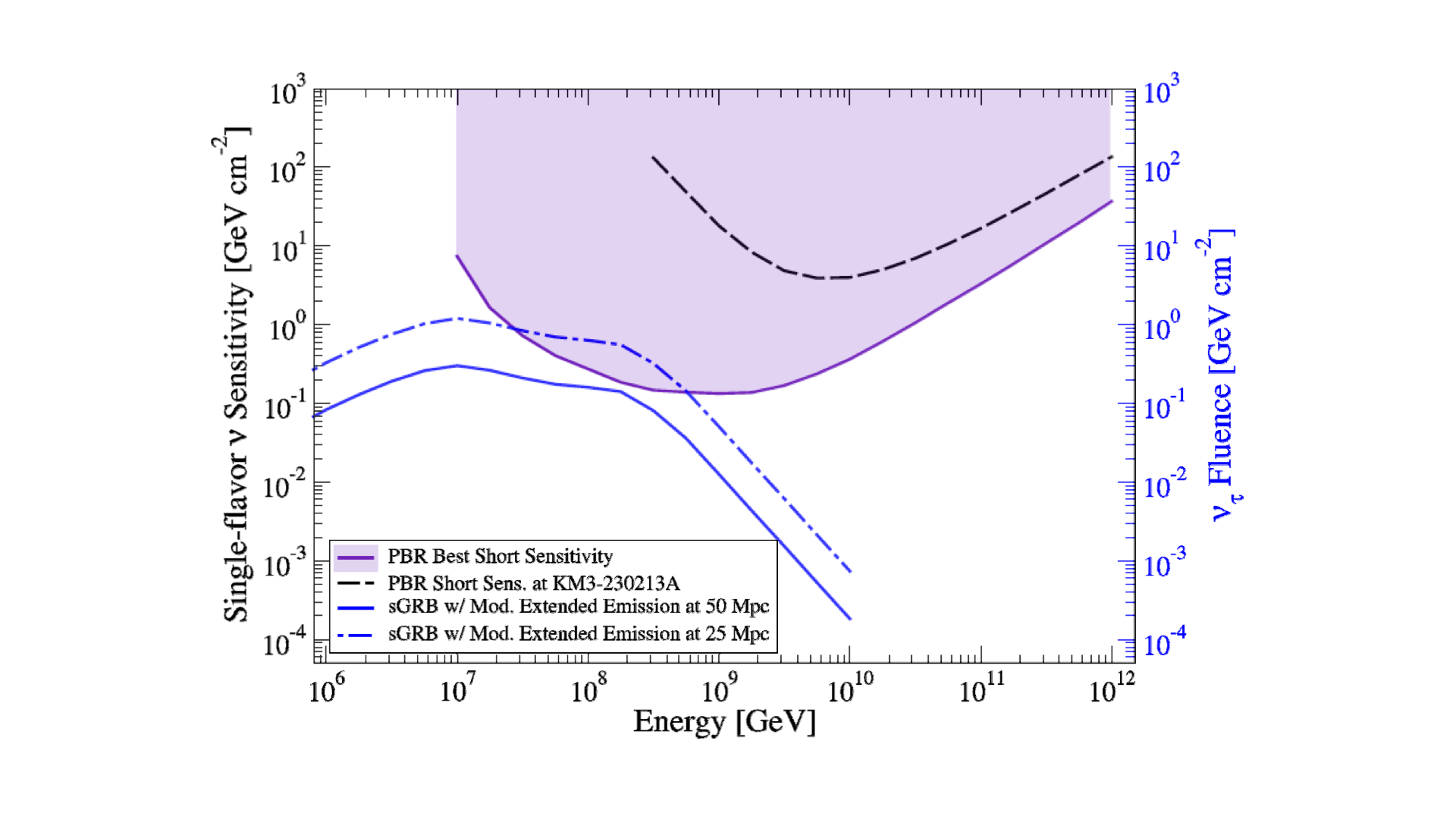} }}%
    \caption{Left: The single-flavor sensitivity range for PBR for a 20-day flight launched in early April 2027. For comparison, modeled single-flavor neutrino fluences from a nearby (\unit[3]{Mpc}) binary neutron star (BNS) merger is plotted as blue line \cite{Fang:2017tla}).  Right: Single-flavor short duration transient sentivity for PBR. The dashed and solid lilac lines represent the fluences from a short gamma-ray burst with extended emission \cite{Kimura:2017kan} located at 25 and \unit[50]{Mpc} respectively.}%
    \label{fig:sensitivity_too}%
    \vspace{-2mm}
\end{figure}

To illustrate the potential sensitivity of PBR to long-duration neutrino bursts (lasting $\sim 30$ days), the left panel of \autoref{fig:sensitivity_too} shows the range of projected average sensitivities of PBR ($6.0^\circ \times 12.0^\circ$ CC FoV) for a 20-day flight. For reference, the fluences from binary neutron star mergers at \unit[3]{Mpc} is also shown \cite{Fang:2017tla}. The right panel indicates for the same flight configuration the sensitivity to short bursts lasting \unit[1000]{s}, again the fluences for short GRB with extended emission model (see \cite{Kimura:2017kan} for details) at 25 and \unit[50]{Mpc} are shown.
For a transient neutrino source in the best-case location, the single-flavor sensitivity it shown by the bottom edge of the purple region. Other locations have sensitivities within the purple region (i.e., the location of KM3-230213A \cite{KM3NeT:2025npi} is indicated with dashed black line), which extends beyond the top of the plot since part of the sky is inaccessible to observations. 

\begin{figure}
    \centering
    \includegraphics[width=\textwidth]{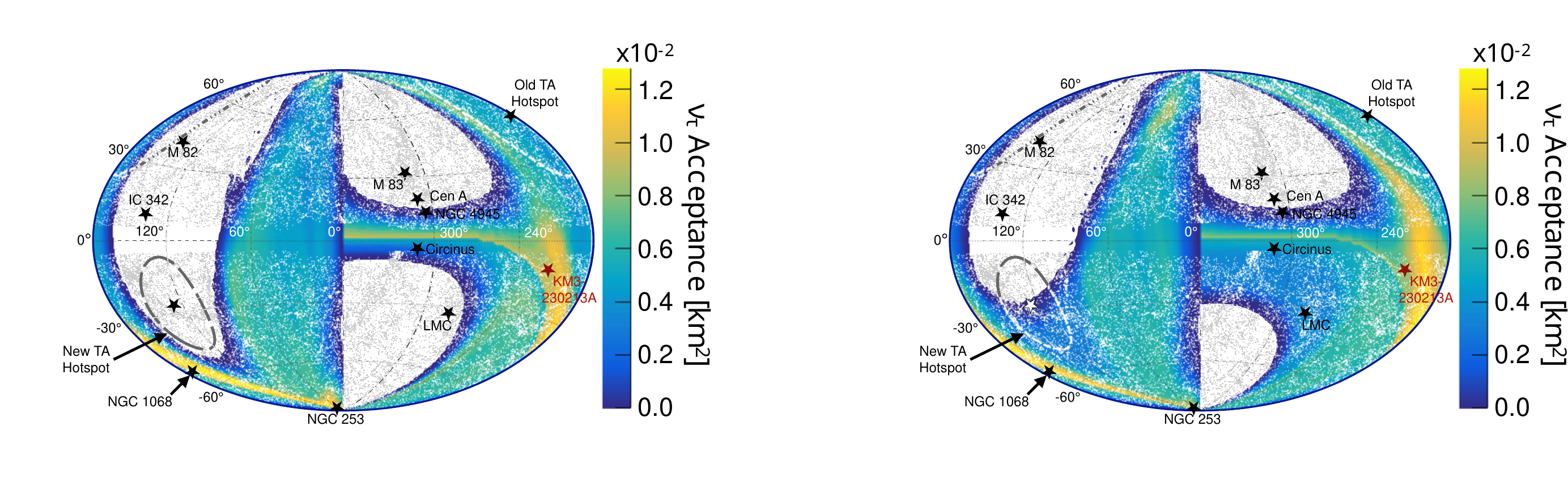}
    \caption{PBR's $\nu_{\tau}$' acceptance assuming a BNS merger model~\cite{Fang:2017tla} at 3 Mpc distance in galactic coordinates. Some point sources are displayed on the map. Nearby sources of interest (shown with black stars, see e.g., \cite{Venters:2019xwi}) and additional sources (grey/white points) from the 2MASS Redshift Survey~\cite{Huchra:2011ii} are included for comparison. The left panel is for a 20 day flight duration, while the right panel is for a mission lasting 50 days. For both plots the balloon trajectory is based on historical wind patterns.}
    \label{fig:skyplot_too}
\end{figure}

The two sky maps in \autoref{fig:skyplot_too} illustrate the time-averaged effective areas for a 20 day flight on the left and for a 50 day flight on the right, assuming a launch in early April 2027 and accounting for the requirements of dark skies during the observation period and assuming historic wind patterns for the ballon trajectory. These acceptances were calculated at \unit[$10^9$] {GeV}. We would like to note that the white regions are areas that are not observable by PBR due to its circulating around the Southern Ocean. The calculations are based on methodology presented in \cite{Venters:2019xwi} and \cite{Reno:2021xos}.
\section{Conclusion}
\label{sec:conclusion}

The PBR experiment represents the next generation of stratospheric balloon payloads: it adapts the space-based POEMMA hybrid‑focal‑surface concept to a 1.1\,m wide‑field Schmidt telescope and integrates a Fluorescence Camera (9216‑pixel MAPMT array, 1\,$\mu$s sampling), a fast Cherenkov Camera (2048‑pixel SiPM array, nanosecond sampling). A Radio Instrument optimized for 60–500\,MHz and a $X/\gamma$-ray detector complement these measurements. Configured for a NASA Super‑Pressure Balloon flight from Wanaka, New Zealand, with a campaign exceeding 20 days in the first half of 2027, PBR validates fluorescence observation from suborbital altitudes, provides the first simultaneous optical and radio Cherenkov measurements of high‑altitude horizontal air showers, and has the potential to provide, for the first time, X-ray measurements possibly shedding light on the earliest shower ages or at least providing crucial background estimation for future missions. Furthermore, PBR executes target‑of‑opportunity follow‑ups to search for Earth‑skimming neutrinos.

By combining both optical techniques in one hybrid focal surface, PBR exercises critical subsystems for POEMMA‑like space missions in a near-space environment. PBR is expected to deliver a large sample of high‑altitude events that can constrain the cosmic‑ray energy spectrum and composition near the PeV scale, provide a unique multi‑hybrid dataset for shower‑physics validation and reconstruction studies. PBR thus serves as a pathfinder that advances near‑term science in the UHECR and VHEN domains while directly informing the design and operational planning of future space observatories.

\acknowledgments
The authors would like to acknowledge the support by NASA award 80NSSC22K1488 and 80NSSC24K1780, by NASA EPSCoR 80NSSC22M0222, by the French space agency CNES and the Italian Space agency ASI. The work is supported by OP JAC financed by ESIF and the MEYS CZ.02.01.01/00/22\_008/0004596. We gratefully acknowledge the collaboration and expert advice provided by the PUEO collaboration. We acknowledge the NASA Balloon Program Office and the Columbia Scientific Balloon Facility and staff for extensive support. We also acknowledge the invaluable contributions of the administrative and technical staffs at our home institutions. This research used resources of the National Energy Research Scientific Computing Center (NERSC), a U.S. Department of Energy Office of Science User Facility operated under Contract No. DE-AC02-05CH11231. The authors thank the company Monolithic Power Systems, Inc. (MPS) in Ettenheim, Germany, in particular Mr. Jan Spindler, for the generous support with EMI measurements in their anechoic chamber.

\bibliographystyle{JHEP}
\bibliography{pbr.bib}

\end{document}